\documentclass{aa}
\def\ltsim{\lower.5ex\hbox{$\; \buildrel < \over \sim \;$}}
\usepackage{apj2aa}
\usepackage{astron} 
\usepackage{float,epsfig,psfig}
\usepackage{pstricks}

\begin{document}
% \thesaurus{11.03.1, 12.03.3, 12.12.1
% }

\title{Temperature gradients in XMM-Newton 
observed REFLEX-DXL galaxy clusters
at $z\sim 0.3$\thanks{This work is based on observations
made with the XMM-Newton, an ESA science mission with 
instruments and contributions directly funded by
ESA member states and the USA (NASA).}}
\author{Y.-Y. Zhang\inst{1}, A. Finoguenov\inst{1}, 
H. B\"ohringer\inst{1},
Y. Ikebe\inst{1}$^{,}$\inst{2}, K. Matsushita\inst{1}$^{,}$\inst{3}, 
and P. Schuecker\inst{1}}

\offprints{Y.-Y. Zhang, \\
e-mail: yyzhang@mpe.mpg.de}

\institute{
Max-Planck-Institut f\"ur extraterrestrische Physik,
Giessenbachstra\ss e, 85748 Garching, Germany
\and Joint Center for Astrophysics, University of Maryland, Baltimore Country, 1000 Hilltop Circle, Baltimore, MD 21250, USA
\and Tokyo University of Science, Tokyo, Japan}

\date{Received 
5 June 2003
 / accepted 22 September 2003}
\authorrunning{Zhang et al.}
\titlerunning{Temperature gradients in XMM-Newton 
observed REFLEX-DXL galaxy clusters at $z\sim 0.3$}

\abstract{
We present XMM-Newton results on the temperature profiles of a
volume-limited sample of galaxy clusters at redshifts $z\sim0.3$,
selected from the REFLEX survey (REFLEX-DXL sample).  
In the spectral analysis, where only the energies above 1~keV were
considered, we obtained consistent results on the temperature derived
from the pn, MOS1 and MOS2 data. Useful temperature measurements
could be performed out to radii with overdensity 500 ($r_{500}$) for
all nine clusters. We discovered a diversity in the temperature
gradients at the outer cluster radii with examples of both flat and
strongly decreasing profiles.  Using the total mass and the gas mass
profiles for the cluster RXCJ0307.0$-$2840 we demonstrate that the
errors on the mass estimates for the REFLEX-DXL clusters are within
25\% up to $r_{500}$.

 }

\maketitle
%\oneandhalfspace
\section{Introduction}

The number density of galaxy clusters probes the cosmic evolution of
large-scale structure (LSS) and thus provides an effective test of
cosmological models.  It is sensitve to the matter density
($\Omega_{\rm m}$) and the amplitude of the cosmic power spectra on
cluster scale ($\sigma_8$) (e.g. Schuecker et al. 2003).  Its
evolution is sensitive to the dark energy ($\Omega_{\Lambda}$)
(e.g. Vikhlinin et al. 2002).  The most massive clusters are
especially important in tracing LSS evolution since they are expected
to show the largest evolutionary effects. In addition, the X-ray
properties of the most massive clusters should be easier to describe
in hierarchical modeling since the structure of the X-ray emitting
intracluster plasma is essentially determined by gravitational effects
and shock heating. With decreasing cluster mass and intracluster
medium (ICM) temperature, non-gravitational effects play an important
role before and after the shock heating (Voit \& Bryan 2001; Voit et
al. 2002; Zhang \& Wu 2003; Ponman et al. 2003).  Therefore, the most
massive clusters provide the cleanest results in comparing theory with
observations.

  \begin{table*}
  {
  \begin{center}
  \footnotesize
  {\renewcommand{\arraystretch}{1.3}
  \caption[]{Compilation of some observational information on the 
nine REFLEX-DXL
clusters. Col. (1): Cluster name. Cols. (2--3):
Sky coordinates. Cols. (4--6): Net
exposure time of MOS1, MOS2 and pn after cleaning for the flaring
episodes. Cols. (7--9): Light curve cleaning upper limit.
Col. (10): Hydrogen column density in units of $10^{20}{\rm
cm}^{-2}$ (Dickey
\& Lockman 1990). Col. (11): Revolution of
XMM-Newton.  Col. (12): Alternative name
}
  \label{t:primarytab}}
  \begin{tabular}{lrrrrrcccccl}
\hline   
\hline     
Cluster & $\alpha$ ($^o$) & $\delta$ ($^o$) & \multicolumn{3}{c}{Exposure Time
(s)} & \multicolumn{3}{c}{Criteria (${\rm cts/100s}$)} & $n_{\rm
H}$ &  orbit & Alternative\\ 
(RXCJ) & \multicolumn{2}{c}{Eq. J2000.0} & MOS1 & MOS2 & pn & MOS1 & MOS2 & pn &
&  & name\\
\hline 
    $ 0014.3-3022$ &   $ 3.5837$ &  $-30.3757$ & 15085 & 15510 & 10057 & 23.4 & 23.7 & 61.2 & 1.60 & 270 & A2744 (AC118)\\
    $ 0043.4-2037$ &  $ 10.8508$ & $ -20.6225$ & 11253 & 11248 & 6318 & 23.5 & 25.0 & 69.8 & 1.54 & 380 & A2813\\
    $ 0232.2-4420$ &  $ 38.0717$ & $ -44.3453$ & 11979 & 11508 &  7741 & 22.2 & 22.4 & 63.4 & 2.49 & 474 \\
   $  0307.0-2840$ &  $ 46.7667$ & $ -28.6708$ & 12309 & 12610 &  8126 & 21.9 & 22.5 & 56.8 & 1.36 & 218 & A3088 \\
   $  0528.9-3927$ & $  82.2342$ & $ -39.4636$ &  7097 &  6806 &  3297 & 23.3 & 23.2 & 57.4 & 2.13 & 324 \\
   $  0532.9-3701$ &  $ 83.2350$ & $ -37.0260$ & 10374 & 11191 & 6527 & 23.3 & 25.3 & 63.0 & 2.90 & 518 \\
   $  0658.5-5556 $  & $ 104.5700$ &$  -55.9600$ & 25339 & 23365 & 18307 & 24.7 & 23.6 & 57.7 & 6.53 & 159 & 1ES0657-558\\
   $  1131.9-1955$ & $ 172.9858$ & $ -19.9258$ & 11660 & 11164 &  8511 & 22.3 & 22.3 & 57.8 & 4.50 & 286 & A1300\\
   $  2337.6+0016$ & $ 354.4204$ &  $  0.2760$ & 12216 & 11915 &  7412 & 22.4 & 23.3 & 69.1 & 3.82 & 365 & A2631\\
\hline  
\hline  
  \end{tabular}
  \end{center}
\hspace*{0.3cm}{\footnotesize 
} } \end{table*}

In this project we are analysing an almost volume-complete sample of thirteen
X-ray luminous ($L_{X}~\geq~10^{45}~{\rm erg~s^{-1}}$ 
for $0.1-2.4~{\rm keV}$)
clusters selected from the ROSAT-ESO Flux-Limited X-ray (REFLEX)
galaxy cluster survey (B\"ohringer et al. 2001a) in the redshift
interval $z=0.27$ to $0.31$. There is only a very small correction
to the volume completeness with a well known selection function
for $L_{X}~\geq~10^{45}~{\rm erg~s^{-1}}$ at the higher redshift  
as described in B\"ohringer et al.  (2003; Paper~I).
With this REFLEX-DXL (Distant X-ray
Luminous) sample we want to obtain a robust measure of the cluster
abundance of this epoch, in particular to perform studies of the
evolution of the cluster population by comparing these observations with more
nearby and more distant clusters. A prime goal is to obtain reliable
ICM temperatures of these clusters as a measure of the cluster masses
(e.g. Evrard 1997). Since peculiarities in the cluster structure introduce a
scatter in the mass--temperature relation and since in particular
on-going cluster mergers can lead to a temporary increase in the
cluster temperature and X-ray luminosity (Randall et al. 2002), we aim
for a detailed study of the deep XMM-Newton observations described
here. The clusters are also scheduled for a detailed spectroscopic
study of the cluster dynamics with the ESO-VLT-VIMOS instrument.

The selection of the REFLEX-DXL sample and its properties are
described in detail in Paper\,I. For all
clusters in this sample the XMM-Newton observations have confirmed
that the X-ray luminosity is dominated by diffuse thermal emission
from the ICM of these systems. Therefore, the REFLEX-DXL sample
contributes a unique sample of X-ray luminous and consequently very
massive clusters from roughly the same epoch, which are not only
interesting as cosmological probes, but also for astrophysical studies
like the statistics of cluster substructure, galaxy evolution,
Sunyaev-Zel'dovich observations and many other applications
(e.g. B\"ohringer et al. 2001b).

The estimate or derivation of the cluster mass is an essential step in
almost all these studies. The mass can be either approximately
estimated from the temperature (Evrard 1997), or determined from the
temperature and density distributions of the ICM under the assumption
of hydrostatic equilibrium of the intracluster gas (e.g. Cavaliere \&
Fusco-Femiano 1976; Serio et al. 1981), or otherwise determined from the
mass of the intracluster gas and the assumption of the universality of
the cluster baryon fraction (e.g. White et al. 1993; Vikhlinin et
al. 2002).  The first two methods require a robust determination of
the ICM temperature and a good understanding of the cluster structure for a
reliable interpretation of the results.

Therefore, it is the aim of this paper to establish a reliable method
of spatially resolved temperature determination for the clusters in
the REFLEX-DXL sample and to derive temperature profiles for all the
clusters.  XMM-Newton with its superior sensitivity combined with its good
spatial resolution provides the best means for such studies (Arnaud et
al. 2002). Previously, large data sets on cluster temperature profiles
have been compiled from ASCA (e.g. Markevitch et al. 1998; White 2000;
Finoguenov et al. 2001a; Finoguenov et al. 2002; Sanderson et al. 2003)
and BeppoSAX observations (Molendi \& De Grandi 1999; Ettori et
al. 2002).

This paper is structured as follows.  In Sect.~\ref{s:method}, we
describe the background components, which are important to this
study. Then we present a double background subtraction method, which
is developed to provide a precise background removal. In
Sect.~\ref{s:result}, we analyse the properties of the hot gas in the
galaxy clusters, and show our analytic temperature model.  Then we
determine the total mass and gas mass fraction in the cluster
RXCJ0307.0$-$2840 based on the precise temperature and gas density
profiles. In Sect.~\ref{s:conclusion}, we draw our conclusions.  We
adopt a flat $\Lambda$CDM cosmology with the density parameter
$\Omega_{\rm m}=0.3$ and the Hubble constant $H_{\rm
0}=70$~km~s$^{-1}$~Mpc$^{-1}$. Error bars correspond to the 68\%
confidence level, unless explicitly stated otherwise.

\section{Method}
\label{s:method}

  \begin{table} { \begin{center} \footnotesize
  {\renewcommand{\arraystretch}{1.3} \caption[]{Parameters of the
  residual background models fitted in the 0.4--15~keV band.  
  Col. (1): Cluster name. Cols. (2--4):
  Index of the ``powerlaw/b'' residual background model for MOS1, MOS2
  and pn.  Cols. (5--7): Normalization {\bf at 1~keV} 
  of the ``powerlaw/b'' residual
  background model scaled to 1~arcmin$^2$ area 
  for MOS1, MOS2 and pn in units of $10 ^{-4}$
  {$\rm cts~s^{-1}~keV^{-1}~arcmin^{-2}$}.} 
  \label{t:powb}}
  \begin{tabular}{lcccccc}
\hline   
\hline     
Cluster       & \multicolumn{3}{c}{Index} &  \multicolumn{3}{c}
%{Normalization (pixel$^{-2}$)} \\ 
{Normalization } \\ 
   (RXCJ)           & \multicolumn{3}{c}{MOS1 MOS2 pn} & 
\multicolumn{3}{c}{MOS1 MOS2 pn}  \\
\hline   
 0014.3$-$3022 &     1.47 &     1.48 &     1.95 & 
 $ 1.21     $  & $ 1.73     $  & $ 6.49     $  \\
 0043.4$-$2037 &     1.71 &     1.43 &     1.61 & 
 $ 1.57     $  & $ 1.28     $  & $ 8.50     $  \\
 0232.2$-$4420 &     1.26 &     1.52 &     1.54 & 
 $ 1.26     $  & $ 1.73     $  & $ 6.60     $   \\
 0307.0$-$2840 &     1.28 &     1.43 &  1.82 &  
 $ 0.89     $  & $ 1.21     $  & $ 2.80    $ \\
 0528.9$-$3927 &     0.80 &     0.96 &     1.56 & 
 $ 0.93     $  & $ 1.12     $  & $ 3.36     $ \\
 0532.9$-$3701 &     1.60 &     1.67 &     2.08 & 
 $ 0.32     $  & $ 1.51     $  & $ 3.89     $  \\
 0658.5$-$5556   &     1.52 &     1.64 &     1.95 & 
 $ 7.66     $  & $ 3.99     $  & $ 15.26     $  \\
 1131.9$-$1955 &     1.98 &     2.34 &     3.19 & 
 $ 1.73     $  & $ 1.89     $  & $ 5.96     $  \\
 2337.6$+$0016 &     1.24 &     1.47 &     1.43 & 
 $ 1.71     $  & $ 1.58     $  & $ 8.74     $   \\
\hline  
\hline  
  \end{tabular}
  \end{center}
\hspace*{0.3cm}{\footnotesize 
}
  }
  \end{table}

\subsection{Data preparation}

Of the thirteen XMM-Newton observations of REFLEX-DXL clusters
conducted so far, eight have sufficient quality for the detailed
studies described here. The remaining five clusters are heavily
affected by soft proton flares. Some properties of these observations
are described in Paper\,I.  Re-observations of these targets have been
allocated. An additional X-ray luminous
REFLEX cluster RXCJ0307.0$-$2840 at $z=0.2578$ was observed in this
project and is also analysed here.  It has very good observational
data and appears to be a very regular and symmetric cluster. We have
therefore selected this object as an example to demonstrate our method
of analysis.
 
An overview on the observational data of the complete sample of
thirteen plus one clusters is given in Paper\,I. In this paper we 
compiled further observational information on the sample targets in
Table~\ref{t:primarytab}, which includes 
the observational parameters of the data 
and the alternative names of these targets.

We use the XMMSAS v5.4 software for data reduction.  The MOS and pn
data were taken in standard Full Frame mode and Extended Full Frame
mode, respectively.  For all detectors, the thin filter has been used.

\begin{figure}
\begin{center}
\includegraphics[bb=78 43 570 760,width=6.4cm,angle=-90,clip]
{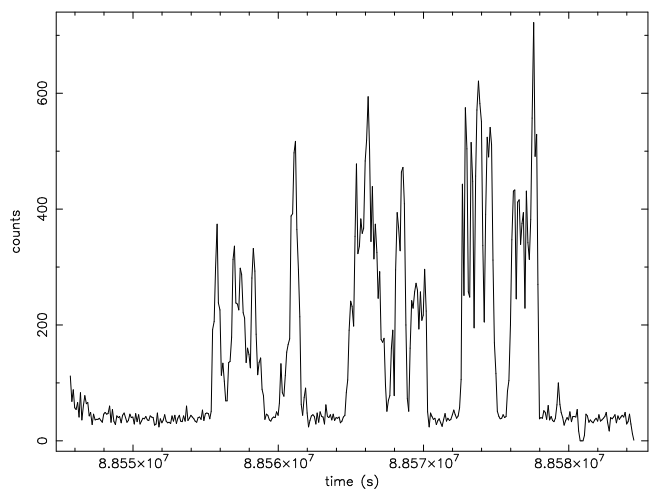}
\end{center}
\figcaption{pn light curve of RXCJ0658.5$-$5556 in the  
10--15~keV energy band. 
{\bf Time is measured in second relative to the XMM-Newton 
internal clock.}
\label{f:lc}
}
\end{figure}

Above 10~keV, there is little X-ray emission from the clusters due to
the low telescope efficiency at these energies; the particle
background therefore completely dominates.  The cluster emission is
not variable, so any spectral range can be used for temporal variability
studies of the background.  
Therefore, the 10--15~keV energy band (binned in 100~s
intervals) was used to monitor the particle background and to excise
periods of high particle flux. In this screening process we use the
settings $FLAG=0$ and $PATTERN<5$ for pn, and $PATTERN<13$ for MOS.
As an example, Fig.~\ref{f:lc} shows the 10--15~keV pn
light curve of RXCJ0658.5$-$5556.

\goodbreak
\goodbreak

We reject those time intervals affected by flares 
in which the detector countrate (ctr)
exceeds a threshold of $2 \sigma$ above the average ctr,
where the average and the variance have been interactively determined 
from the ctr histogram below the rejection threshold.
A similar cleaning criterion is applied to
the screening of the background observation. We note, however, that
the thresholds will be different for the source and background
accumulations. Formal freezing of the cleaning criterion does not
overcome the difference in the mean background ctr.  In our
analysis we searched through a number of background observations to
find the one matching our target observations.  The selection
criterion is therefore to find the one with a similar cleaning
threshold.

\goodbreak
\begin{figure}[!ht]
\begin{center}
\includegraphics[width=7.cm]{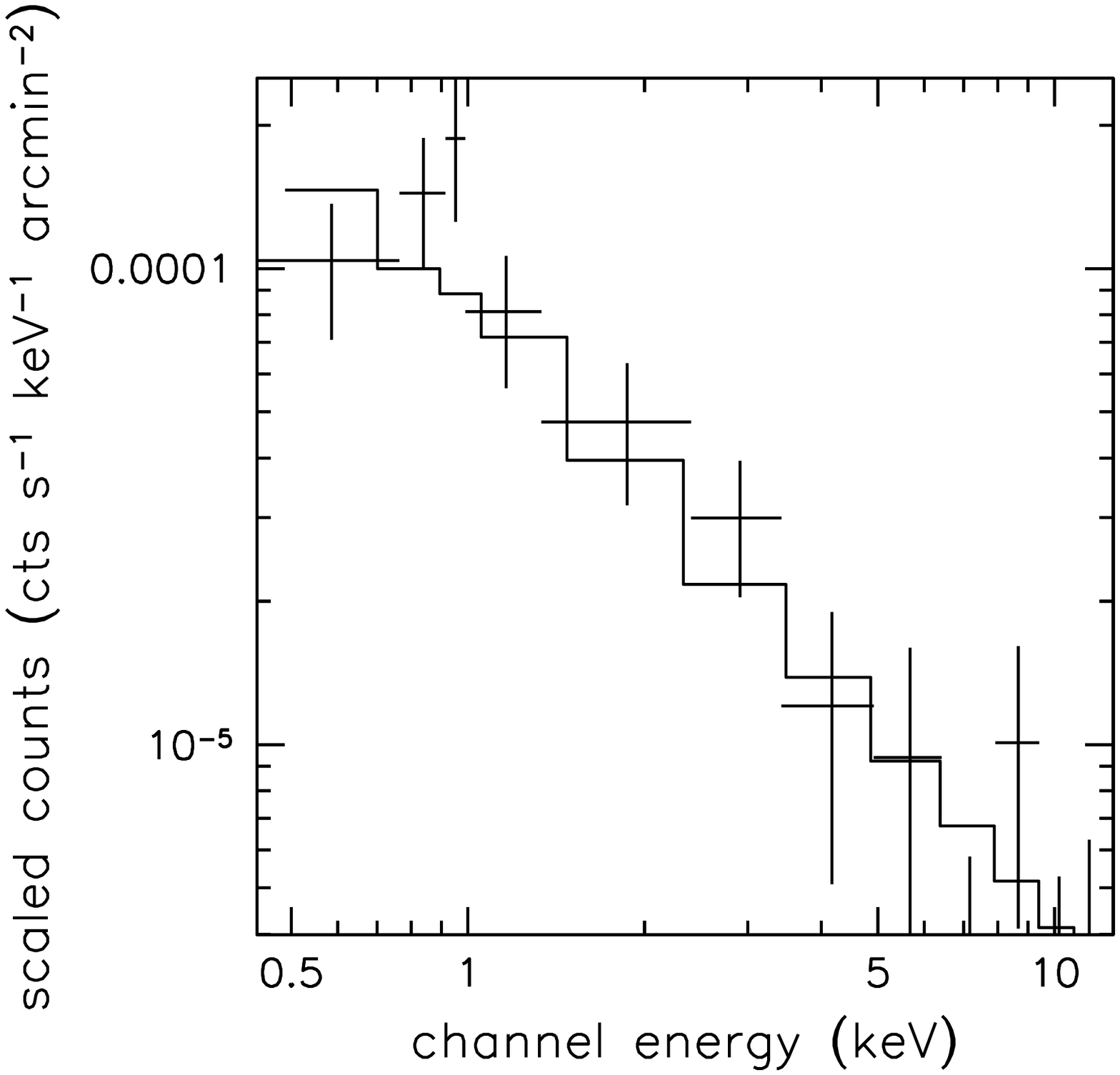}
\includegraphics[width=7.cm]{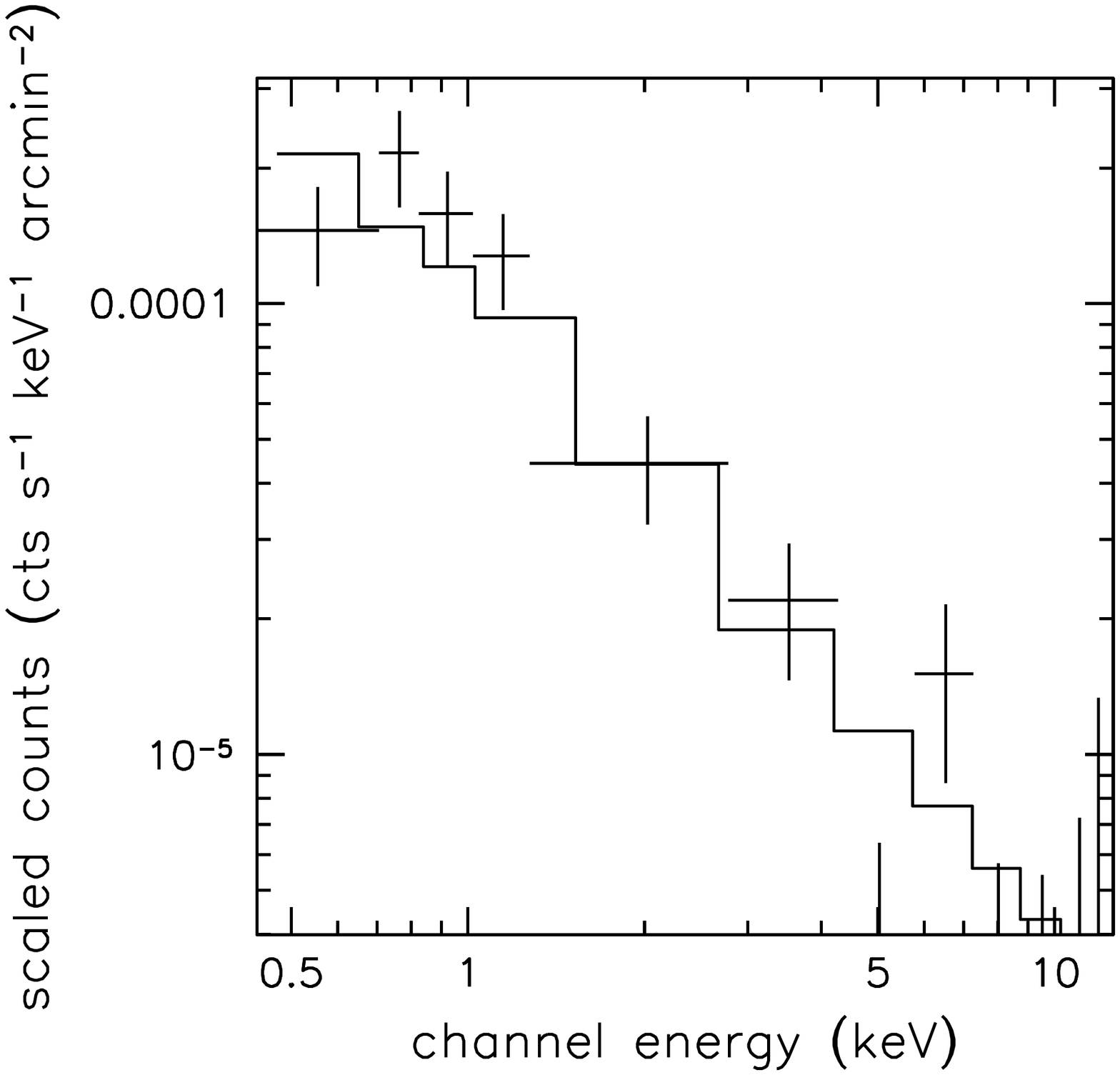}
\includegraphics[width=7.cm]{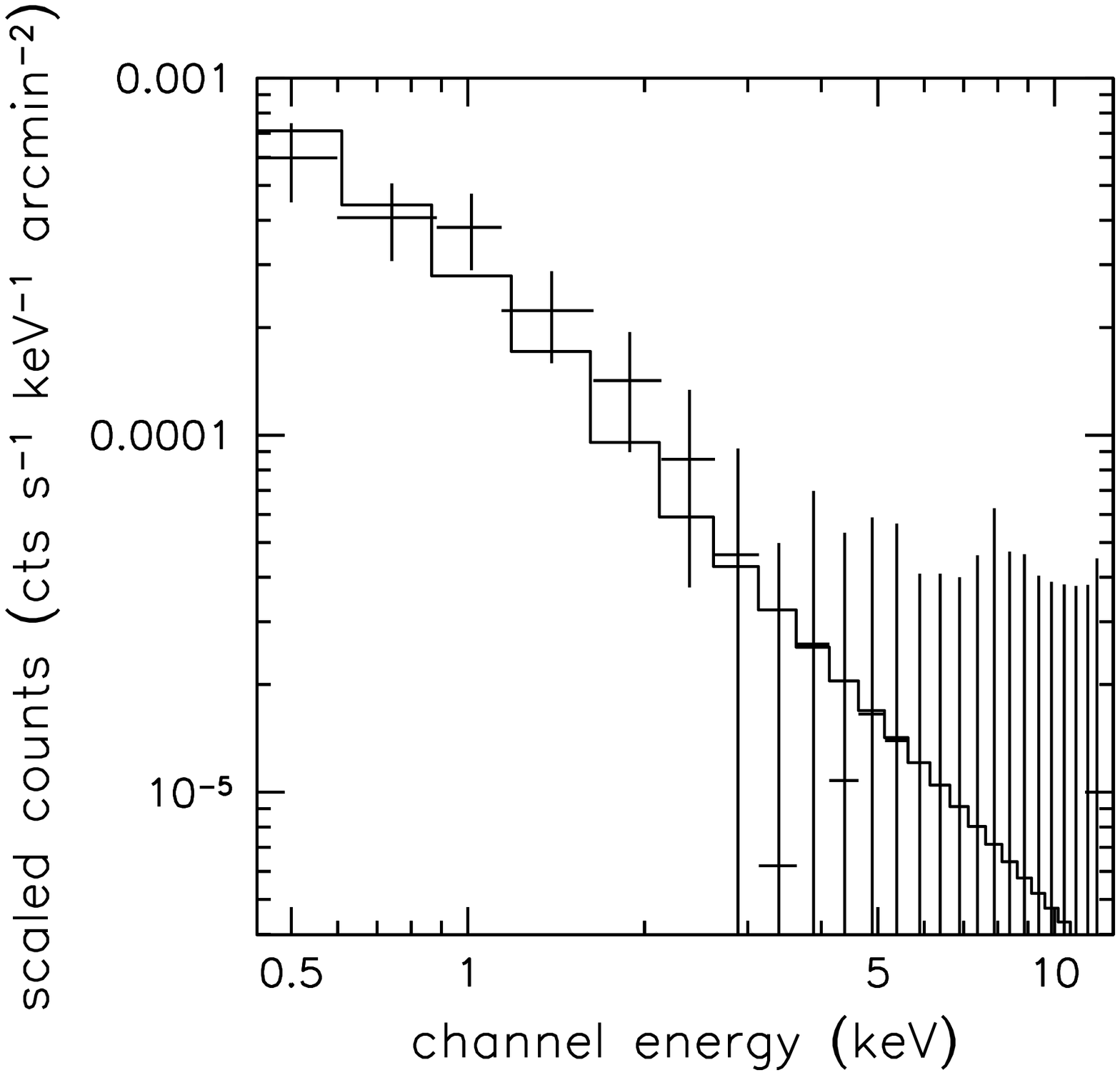}
\end{center}
\figcaption{
%``Powerlaw/b'' fits (histogram) of the 
Residual background (after subtraction of the background 
obtained from the XMM-Newton observations of CDFS) of RXCJ0307.0$-$2840
for the 9.2--11.5$^\prime$ region from the pointing centers of the 
MOS1 (top), MOS2 (middle) and pn (bottom) 
scaled to 1~arcmin$^2$ area. The data are fitted 
by a power law model (solid lines).
\label{f:edgefit}
}
\end{figure}
\goodbreak

As shown in Table~\ref{t:primarytab}, the pn and MOS detectors 
all have their own similar
cleaning thresholds for all observations.  
In Sect.~\ref{s:mos} we consider in detail how this background behaviour
affects the temperature determination. 

In the analysis of the pn data, we statistically remove the out-of-time
effect by creating an out-of-time (OOT) event list file and using the 
XMMSAS products such as images and spectra to subtract it. 
The present observations have been taken 
in the Extended Full Frame mode (frame~time=199~ms). {\bf At this mode,
the fraction of OOT effect amounts 2.32\%, which we used to normalize 
the XMMSAS OOT product before we subtract it from the  
XMMSAS normal product.} 

\subsection{Background analysis}
\label{s:mos}

The purpose of the background analysis described in the following 
is to find a suitable ``blank field'' observation to be used for 
the background subtraction and to further analyse the difference 
between the target and background to take this residual 
background into account.

The background has several components which exhibit different spectral
and temporal characteristics.  In the low energy band ($<0.3$~keV),
the instrumental background is dominated by electronic noise
consisting of a large number of small amplitude events added up during
each frame accumulation (Read 2002).  This noise depends on the
read-out frequency of the cameras and is sensitive to the energy
offset of each individual pixel. Energetic particles produce several
line and continuum components in the background, which can be further
subdivided into time variable and constant components. The
constant component has been intensively studied by Lumb et al. (2002)
and Read \& Ponman (2003). This component can be removed using the so
called blank field observations. De Luca \& Molendi (2001) report some
evidence for a secular evolution of the background level on a half
year time scale. Therefore, background accumulations close to the date
of the target observation are more suitable. In addition, variations
in the instrumental background on a much shorter time scale have been
seen (e.g. De Luca \& Molendi 2001). Part of such periods with an
increased background are rejected through the analysis of the light
curve (e.g. Read \& Ponman 2003). However, we sometimes still observe
a residual enhancement of the background associated with an increase
in the quiet flux of soft protons.  The typical time-scale for the
variation of this `quiet' component is comparable to or exceeds the
typical observational time-scale. Therefore, it is in general not
possible to remove observational intervals affected by this
enhancement.  Chen et al. (2003) describe an example of such an
observation where quite different `quiet' background levels are observed
before and after a flare, respectively.

The photon background consists of foreground emission from the Galaxy
as well as the Cosmic X-ray Background (CXB).  Observations of the
blank field also contain both components, provided the accumulations
are done with the same instrumental set-up (e.g. with a particular
filter) and the spectra of the X-ray background are the same for both
the target and the blank field.  This is only guaranteed for the CXB
and the emission from the Galaxy halo. Since the Galaxy also plays a
role as absorber and foreground emitter, it is important to have a
similar absorbing column density for both the target and background
accumulations. Additionally, there are some extra Galactic components,
that display spatial variations. To constrain them, one has to look into the
ROSAT All-Sky Survey data around both the background accumulations and
the target. Observations with normal conditions of the Galactic
emission are referred to the quiet Galactic zones.  So are most of the
X-ray background accumulations (Lumb et al. 2002; Read \& Ponman
2003). Some source removal is performed on the existing background
accumulations. This changes the shape and intensity of the residual
CXB.  Therefore, a similar source removal has to be performed in the analysis
of the target.

Several available XMM-Newton pointings have been investigated. We
conclude that the XMM-Newton pointings in the Chandra Deep Field South
(CDFS) have similar background conditions as most targets and
sufficient exposure time.  Therefore, the CDFS is a good candidate for
the background for our study. However, there is still a small
difference between the background of our sources and the CDFS,
e.g. the background ctr in the target observations is slightly 
higher {\bf (10--20\%)} than in the CDFS, which
we ascribe mostly to the contamination by soft protons.

\begin{figure}
\begin{center}
\includegraphics[width=7.cm]{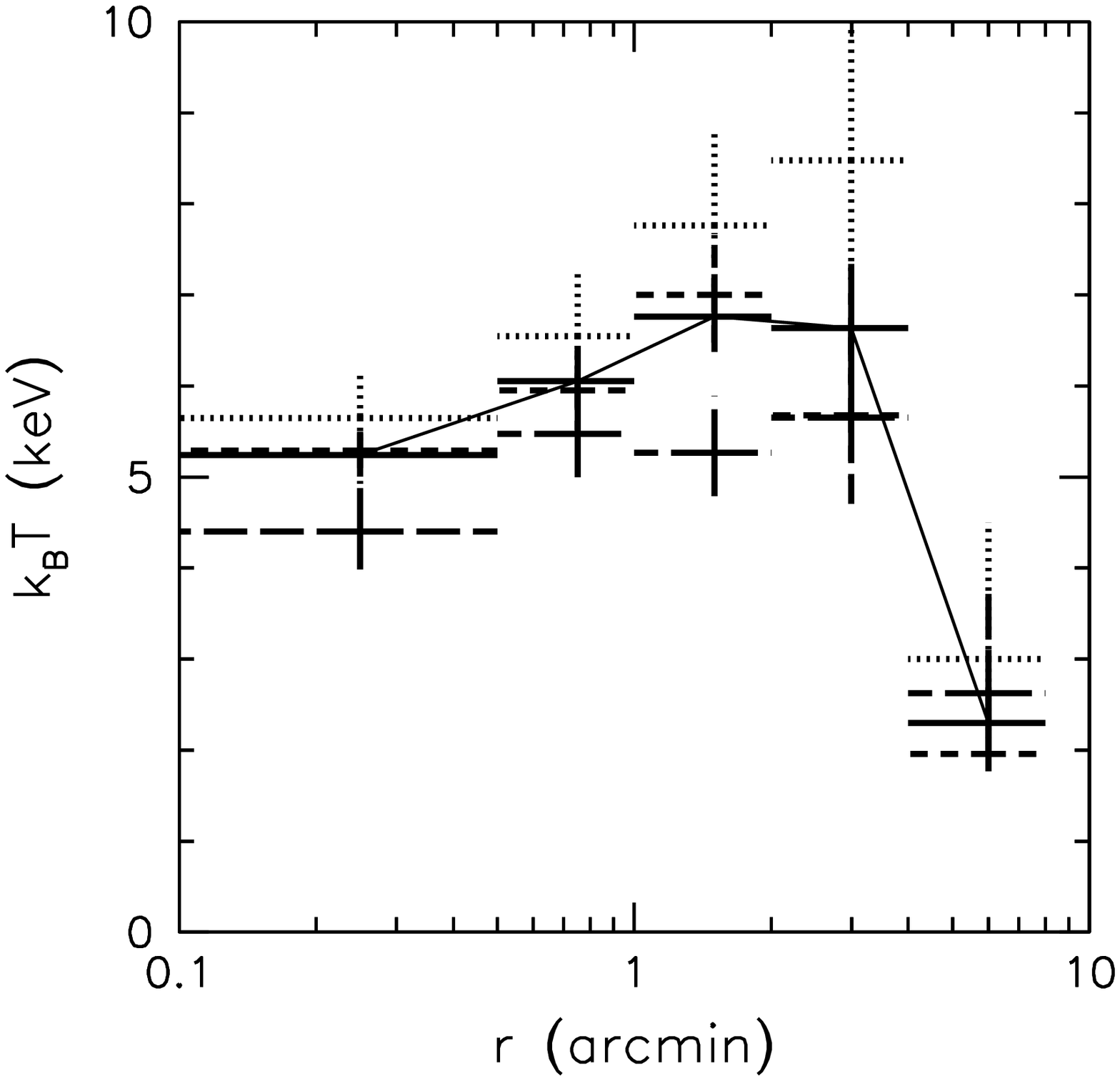}
\includegraphics[width=7.cm]{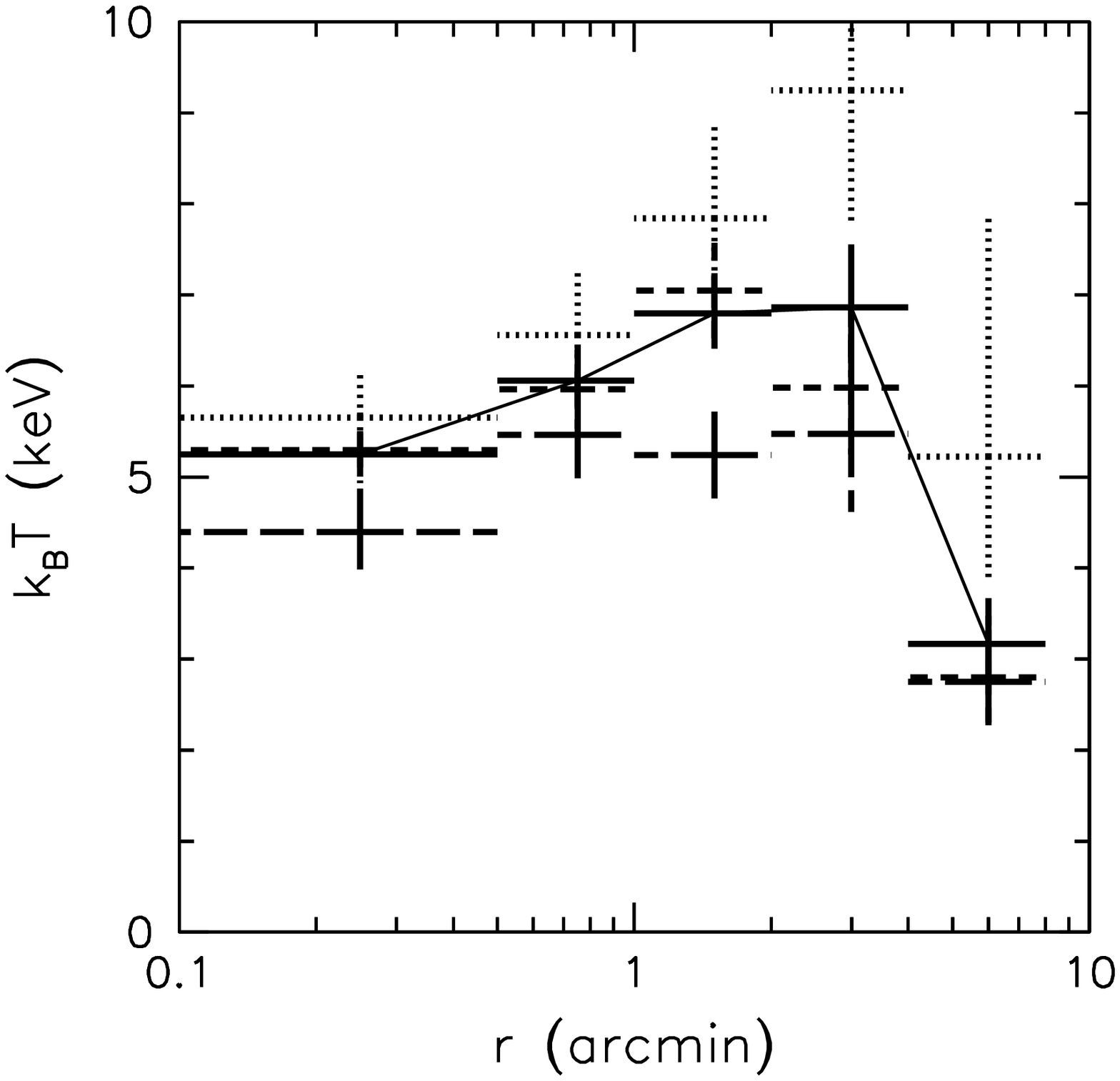}
\figcaption{Temperature profiles for RXCJ0307.0$-$2840 
with residual background subtraction (upper panel) and without 
residual background subtraction (lower panel) fitted in the
0.4--10~keV band for MOS1 data (dotted lines), MOS2 data (dashed
lines), pn data (dash-dotted lines) and combined data (solid
lines). 
%The lower panel shows the case with residual background
%subtraction but fitted in the 1--10~keV band.  
Additional solid lines
connect the temperature measurements for the combined data.
\label{f:ktprof}
}
\end{center}
\end{figure}

\begin{figure}
\begin{center}
\includegraphics[width=7.cm]{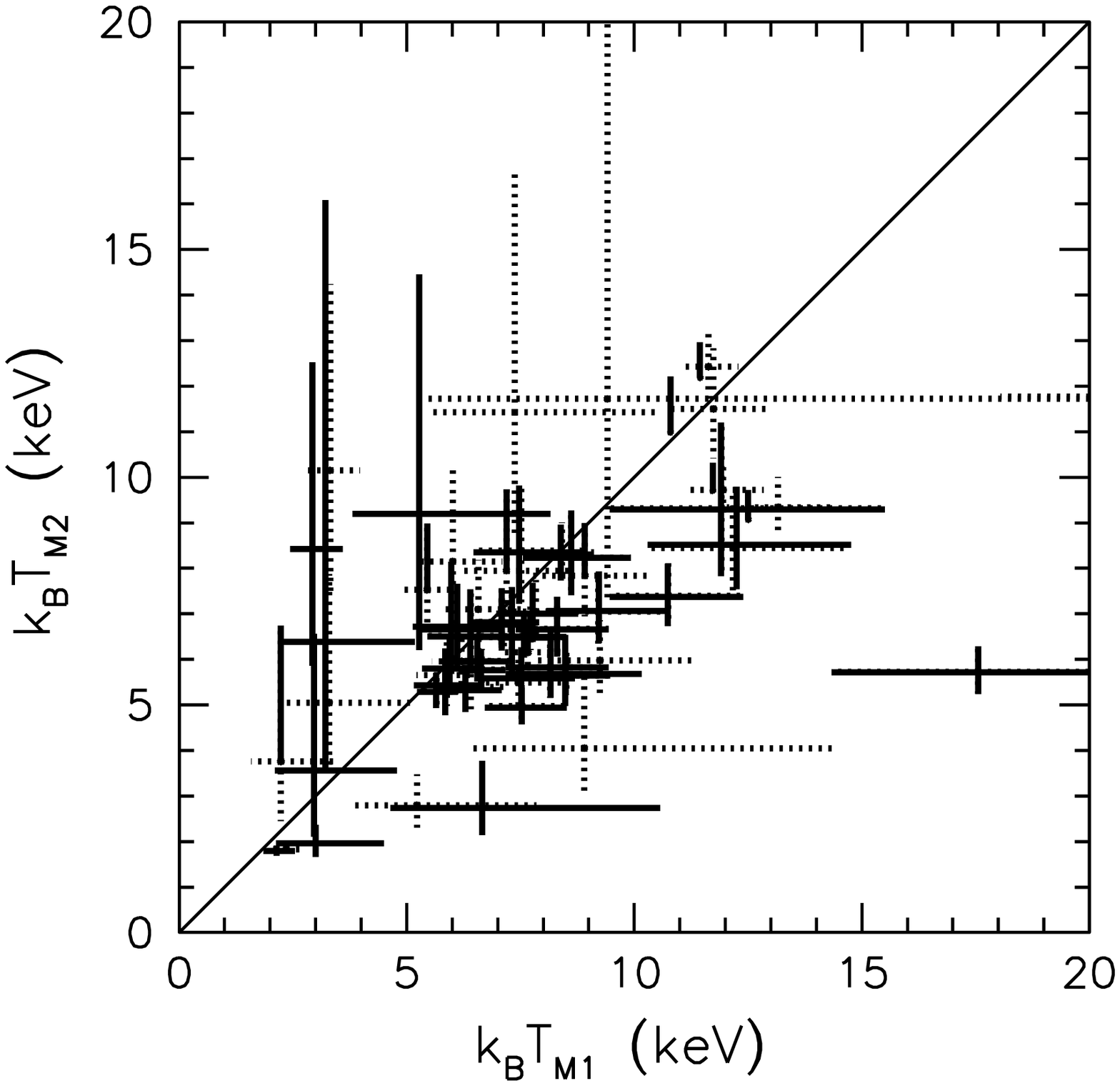}
\includegraphics[width=7.cm]{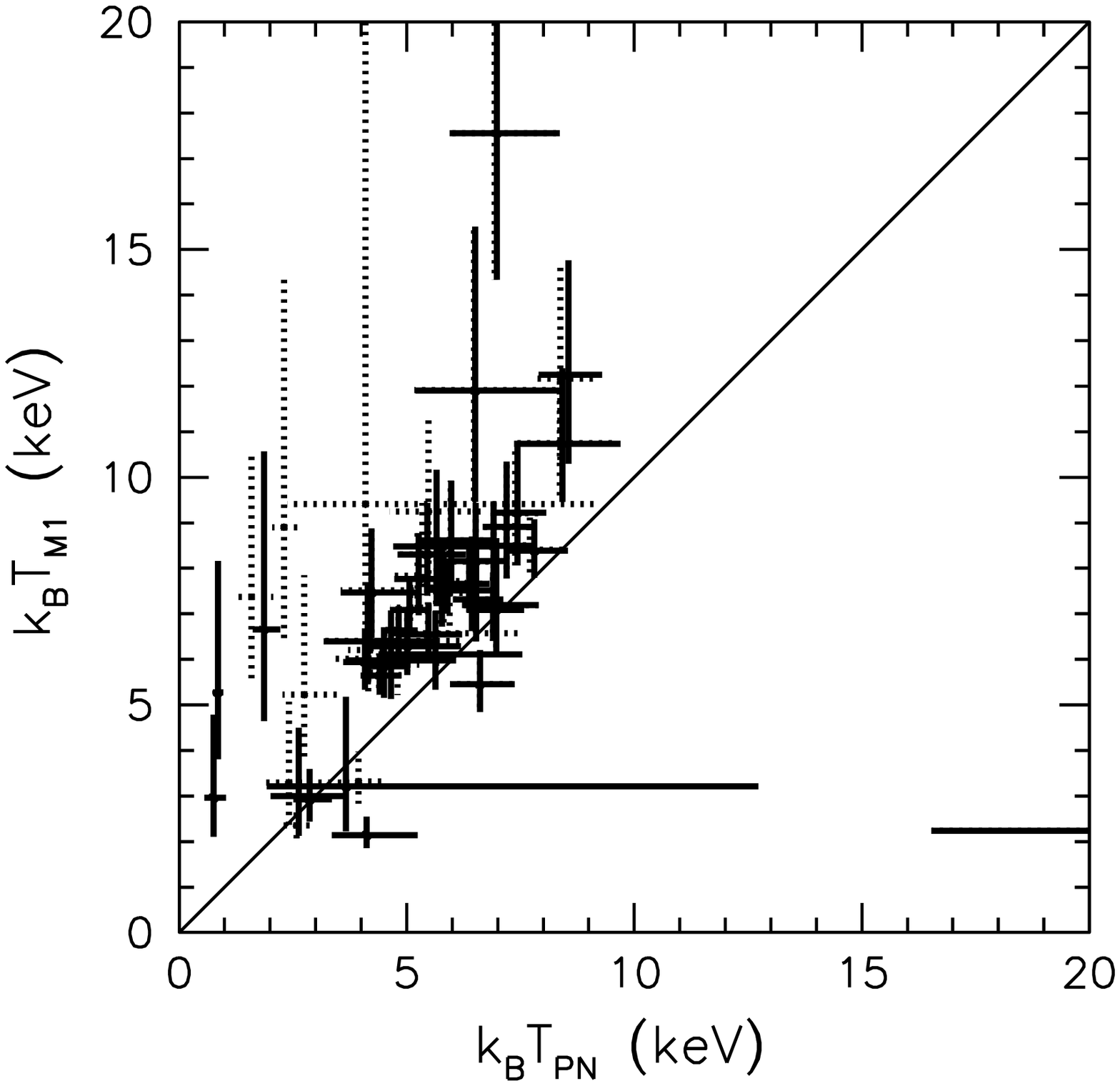}
\includegraphics[width=7.cm]{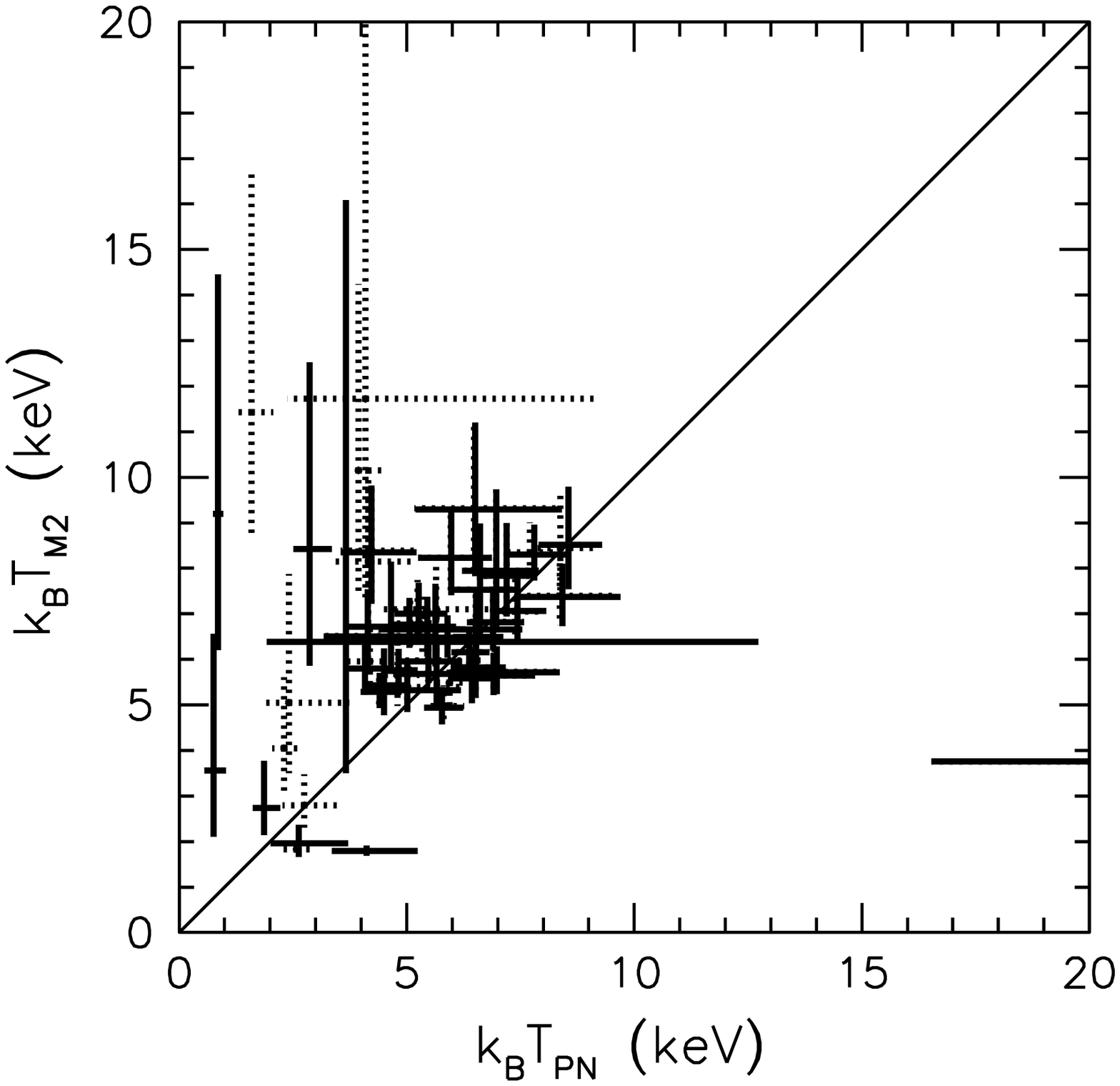}
\figcaption{Comparison of the temperature estimates for 
RXCJ0014.3$-$3022, RXCJ0043.4$-$2037, 
RXCJ0232.2$-$4420, RXCJ0307.0$-$2840, RXCJ0528.9$-$3927,
RXCJ0532.9$-$3701, RXCJ0658.5$-$5556, RXCJ1131.9$-$1955 
and RXCJ2337.6$+$0016
from MOS1, MOS2 and pn fitted in the 0.4--10~keV band with 
residual background subtraction (solid
lines) and without residual background subtraction (dotted lines).
\label{f:ktcom}}
\end{center}
\end{figure}

We have carefully planned this cluster study such that the
radii in which the cluster emission can be observed 
extent up to spherical overdensities of 500, i.e., the ratio of
the mean density of the dark halo with 
respect to the redshift-dependent critical density
$\rho_{\rm crit}(z)$. This is the region to which the cluster X-ray emission is
expected to be essentially confined, covering about half of the field
of view (FOV) of the XMM-Newton telescope.  This enables us to extract
a source spectrum from the background region of the target field for
comparison with a background spectrum from the background field
extracted with the same detector coordinates.  Our residual background
modeling procedure consists of analysing such regions not affected by
cluster emission. We assume little or no vignetting of the soft proton
induced background, as suggested by recent studies (Lumb et al. 2002).
Spectra are extracted from the 9.2--11.5$^\prime$ region from the
pointing center for our source observations and background candidate
observations.  In a first step, we compare the spectra extracted from
the outer regions in the 0.4--15~keV band between the sources and the
background pointings. The residual background signal found after the
subtraction of the background field from the target field in this
outer area is then modeled by a power law spectrum (model ``powerlaw/b''
in XSPEC, a power law background model which is convolved
with the instrumental redistribution matrix but not with the effective
area). We use this model to account for the excess soft
proton background in our observations as compared to the background
field.  This residual power law background model is introduced over
the whole energy range (``wabs$*$mekal$+$powerlaw/b'',
an emission spectrum from 
hot diffuse gas, e.g. Mewe et al. 1985, 
considering the Galactic absorption and modeling
the residual background by a power law), 
which yields the correct shape of
the background component after
the combination with the blank field background (double background subtraction
method). During this procedure, the normalization of the residual
background component is always scaled to the area of the source
extraction region.

In Fig.~\ref{f:edgefit} we present examples of the residual
background. The parameters of the residual background models 
fitted in the 0.4--15~keV band for the
clusters are given in Table~\ref{t:powb}.   
The residual background in pn is higher than in 
MOS because the larger thickness of the pn-pixels leads
to a higher sensitivity to the particle flux.
Since the subtraction of the residual background is only a second 
order correction to the data and because of the large noise in the residual 
background data, we are not attempting a perfect model fitting, but 
approximating the data by a simple power law model. The uncertainty 
in the normalization is, anyway, within 5\% and 10\% for MOS and pn, 
respectively. The correction due to the residual background makes only a 
1--4\% effect in the cluster signals and an up to 10\% effect to the 
temperature determinations for cluster radii $r \leq 4^\prime$. 
At larger radii the effect of the correct background effect is large 
as shown in Fig.~\ref{f:ktprof}, but the uncertainty in the 
approximation of the residual 
background -- a third order effect -- is still small.

To recover the correct spectral shape and normalization of the cluster
component, we need both the response matrix file (rmf) and auxiliary
response file (arf). 
{\bf The following need to be taken into account 
in either rmf or arf: 
(i) Pure redistribution matrix giving the probability that a photon
of energy E, once detected, will be measured in data channel PI.
(ii) Quantum efficiency (with closed filter position) 
of the CCD detector.
(iii) Filter transmission.
(iv) Geometric factors
such as bad pixel corrections and gap corrections (around 4\%).
(v) Telescope effective area as a function of photon energy.
(vi) Vignetting correction to effective area for off-axis pointings.
%(vii) Point spread function corrections.
We choose the rmf which corresponds to 
(i) and (ii) (with closed filter position).} 
The pn rmf that corresponds to this choice has a `\_closed' keyword 
in their naming conventions. For the MOS detectors
we use software to generate such files, kindly provided by S.
Sembay. {\bf The arf corresponds to (iii), (iv) and (v) and is made 
according to the average flux 
detected in the different extraction annuli, which takes (vi) 
into account.} It
is created using the XMMSAS based program `clarf' by A.
Finoguenov. {\bf Furthermore, in our analysis we apply the 
logarithmically spaced radial bins, which provide nearly the same flux per
bin. The importance of the scaterring due to the point-spread-function
(PSF) is therefore suppressed.
Using the XMM-Newton PSF calibrations by Ghizzardi (2001) we have
estimated the loss fraction of the flux: 20\% for the central bins   
($0.5^{\prime}$) and less than 
10\% for the other bins ($\ge 1^{\prime}$) with
energy dependent effects being negligible.}

In summary the spectral analysis is performed in two steps.
(i) A model for the residual background (background
difference) is obtained in XSPEC from a comparison 
of the outer region of the target and background fields 
(see Fig.~\ref{f:edgefit} and Table~\ref{t:powb}).
(ii) The spectral modeling is performed in XSPEC with the cluster 
region as source data, the CDFS as background and the residual 
background as a second model component with model parameters
fixed to the values found in step (i) (see Fig.~\ref{f:zpha0540}). 

In Fig.~\ref{f:ktprof} we present the temperature profiles resulting
from this background subtraction method fitted in the 0.4-10~keV band
for the example object,
RXCJ0307.0$-$2840.  We actually compare the results of the two step
background subtraction considering the residual background 
with the simple one step background
subtraction.  
In Fig.~\ref{f:ktcom} we provide the comparisons of the
results from MOS1, MOS2 and pn fitted in the 0.4--10~keV band. 
One notes that the residual background subtraction
provides results in which all three instruments tend to show a slightly 
better consistency. The upper limit of the uncertainties 
between the instruments goes down from 90\% to 15\% after the residual 
background subtraction. A
detailed treatment of the background does not completely remove the
differences between the instruments. We found systematically lower
temperatures obtained with pn compared to MOS1 and MOS2
(partially). Because the pn detector is sensitive to the soft component, 
the pn measurements are easily affected by the 
soft band, which results in the lower temperatures given by pn 
compared to MOS. 
Since the temperature estimates of the pn are more strongly
dependent on the soft energy band compared to MOS, we have carried
out a spectral analysis in the 1--10~keV band. 
%with the results shown in Fig.~\ref{f:ktprof}. 
Despite the larger error bars, all temperature
determinations in the central three bins of the cluster become higher,
once the 0.4--1~keV band is excluded. In the following we will
systematically investigate the effect of an energy band selection in
the temperature and mass estimate.

\section{Results}
\label{s:result}

\subsection{Redshift, mean temperature, and metallicity}

\begin{figure}
\begin{center}
\includegraphics[width=8.5cm]
%[bb=79 43 570 750,width=6cm,angle=-90,clip]
%{plots/0478_zpha_xspec_mewapo4800_4200_bw.ps}
{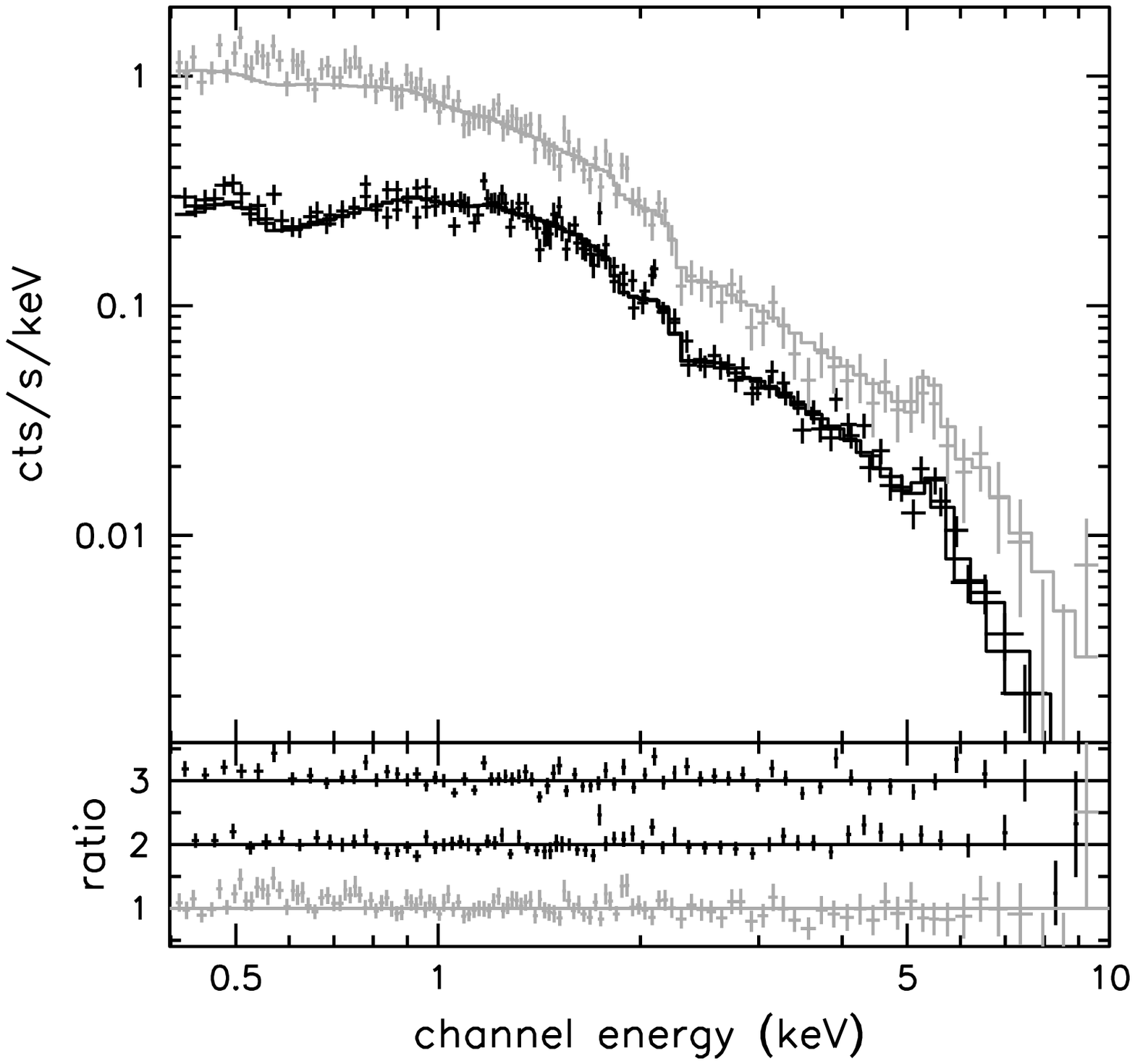}
\includegraphics[width=8.5cm]
{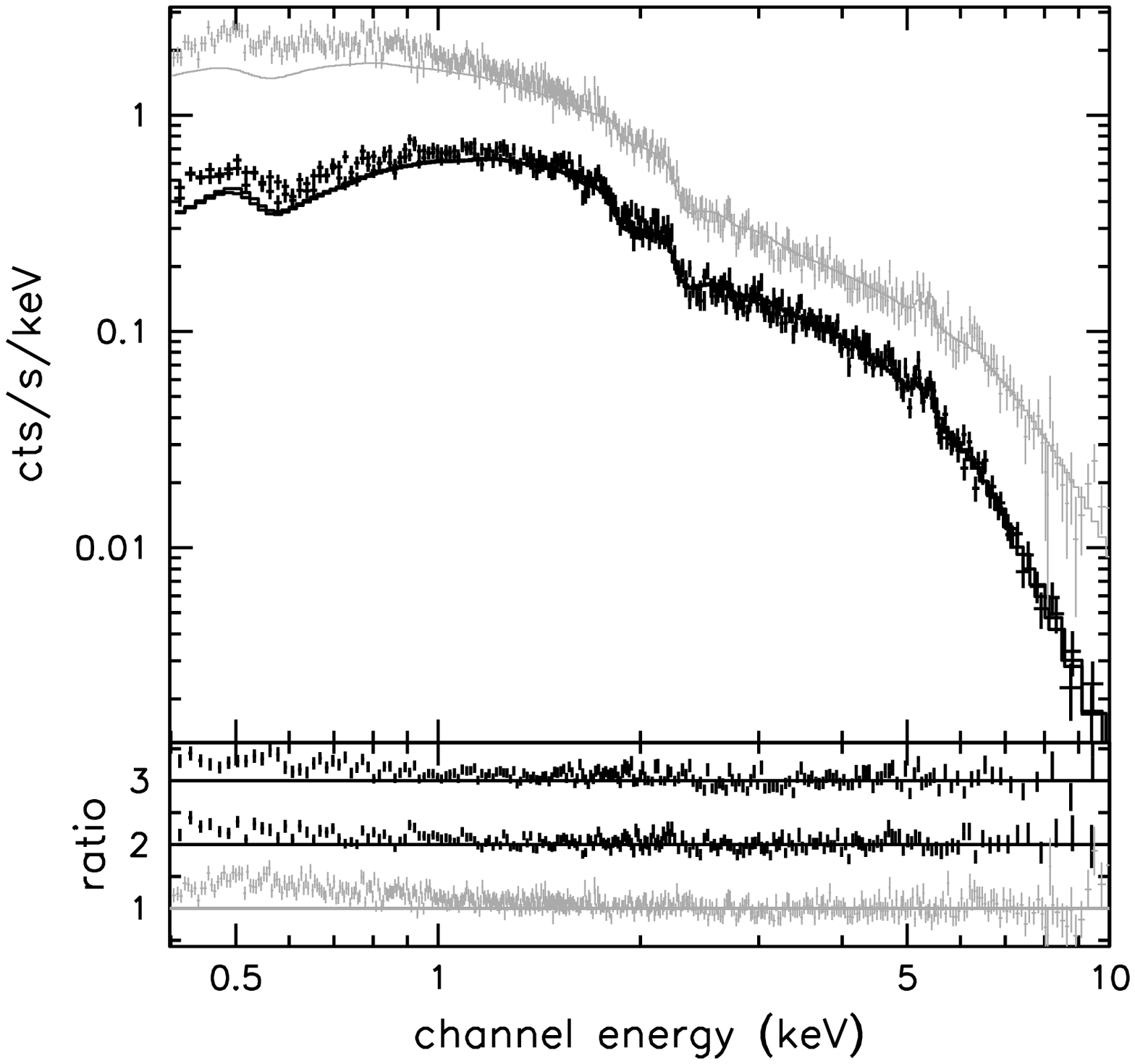}
\end{center}
\figcaption{XMM-Newton spectra of RXCJ0307.0$-$2840 
(top panel, fitted in the 1--10~keV band) 
and RXCJ0658.5-5556 (bottom panel, fitted in the 2--12~keV band) extracted 
from the $0.5<r<4^\prime$ region (pn in grey and MOS
in black) with mekal model, considering the Galactic absorption and modeling
the residual background by a power law.
The ratios of the observational data to the models are in the lower parts
of the panels (offset zero for pn, $+1$ for MOS1, $+2$ for MOS2).
\label{f:zpha0540}
}
\end{figure}

In a first step of the data analysis we derive global properties of
all nine galaxy clusters with good XMM-Newton data. 
Since a fraction of the clusters have cooling
cores, dense central regions with lower temperature and so-called
cooling flows, while the others do not display this phenomenon, we
derive global temperatures including and excluding these regions. In
addition the signal-to-noise decreases fast in the outer regions of
the clusters. Therefore, the global temperature was determined in the
$r<8^\prime$ region and alternatively in the 
$0.5<r<4^\prime$ region (see Fig.~\ref{f:zpha0540}). 
The global temperatures determined in both regions
show some differences. The explanation is partially revealed by the
temperature profiles.  The metallicities in both zones are very
similar.

After subtracting the background and applying the rmf and arf, we fit
the spectra in XSPEC using the one and two step correction models 
(``mekal$*$wabs'', ``mekal$*$wabs$+$powerlaw/b'').  
The fit using the latter model is better.
For the regions covering radii of $0.5<r<4^\prime$ and $r<8^\prime$, 
respectively, 
we use the 0.4--10~keV and
1--10~keV energy bands. Furthermore, we exclude the regions of 
substructure and/or some very bright point sources for several
clusters throughout the procedure (cf. Table~\ref{t:region}).  The
temperature of the small and large substructures in the well-known
post merger cluster RXCJ0658.5$-$5556 are $8.3^{+2.1}_{-1.4}$~keV and
$15.0^{+2.3}_{-1.9}$~keV from the double background subtraction method
in the 2--12~keV band.  We exclude the small substructure to measure
the global temperature. 
The flux within the region we excluded contributes 60\% 
(10$^{\prime \prime}$) 
to 85\% (30$^{\prime \prime}$) to the total flux of the point source. 
The {\bf temperature} measurements vary within 10\% after the subtraction.

The redshifts obtained from the X-ray data (see Table~\ref{t:cl_all})
which are ascribed to the ICM, are consistent with the redshifts
obtained from optical spectra of individual cluster galaxies (see
B\"ohringer et al. 2003) except for RXCJ2337.6$+$0016. 
The optical redshifts contain a heliocentric 
correction, while the uncertainty in the X-ray determined redshifts is one to 
two magnitude larger than this correction and thus no correction was made.
We believe that the optical redshift of RXCJ2337.6$+$0016 
with 5 coincident cluster 
galaxy redshifts is more reliable and accurate than the X-ray result
at this stage. We plan to obtain further information on this object to resolve 
this discrepancy.
   
The measurements of the global temperatures are summarized in
Table~\ref{t:cl_all}. 
Similar to the measurements for the $0.5<r<4^\prime$ region, 
the measurements for the $r<8^\prime$ region 
in the 0.4--10~keV band are lower than in the 1--10~keV band. 
Therefore, we only 
provide the comparison of the measurements for the $0.5<r<4^\prime$ region
fitted in two bands.
The results obtained for the full and restricted
spatial zones are consistent within 1--2$\sigma$ (formal errors)
within 15\% uncertainties.

The global temperatures obtained from model fits
to the larger spectral range 0.4--10~keV are always lower compared to
the temperatures obtained from 1--10~keV. 
To check if the discrepancy is partially due to  
residual Galactic emission, we undertake the following test.
We extract the spectra from the inner (hereafter ``A'') and 
outer (hereafter ``B'') parts of the background region 
in both the background (hereafter ``bkg'') and target (hereafter ``src'') 
observations. 
If there is some difference in the Galactic emission 
between the background and target observations,
there must be some residual Galactic emission after we subtract 
the spectrum ``$B(src)-B(bkg)$'', scaled by the area of the region A, 
from the spectrum ``$A(src)-A(bkg)$'' because of the vignetting effect 
of the Galactic emission.
We apply a thermal emission spectral model (``apec '')
with all parameters free in XSPEC to fit 
this residual emission. We found that the 
temperature of this component is around 0.2~keV and the redshift is 
close to zero. In the following analysis, we fix the temperature
to 0.2~keV, the abundance to solar abundance and the redshift to zero, and
obtain the normalization of this component.
We re-analyse the global properties of the clusters introducing the 
residual Galactic emission normalized by the area and vignetting effect. 
Since the difference in the temperatures determined in the 0.4--10~keV
and 1--10~keV bands still remains, we report the temperature measurements
setting the normalization of the residual Galactic emission to 
its upper limit (cf. Table~\ref{t:cl_all}). However,
this component only makes a $<1$\,\% 
effect in the cluster signals. It is thus clear 
that the discrepancy is not or not mainly due to the residual Galactic 
emission. This discrepancy will again be
discussed below. 

  \begin{table} { \begin{center} \footnotesize
  {\renewcommand{\arraystretch}{1.3} \caption[]{Regions with
  substructures and point sources excluded from the
  analysis. Col. (1): Cluster name. Col. (2--3): Center of the circle
  in sky coordinates for J2000.0. Col. (4): Radii.  Col. (5): ``Yes''
  if there is an optical point-like counterpart in a digitized optical
  survey (e.g. DSS2).} \label{t:region}} \begin{tabular}{lcccc}
\hline   
\hline     
Cluster (RXCJ) & $\alpha$ ($^o$) & $\delta$ ($^o$) & r ($^{\prime\prime}$) & opt\\ 
\hline 
0014.3$-$3022 & $3.6306 $ & $ -30.3754$ & 15 & Yes\\
                & $3.5198 $ & $ -30.4154$ & 15 & Yes\\
                & $3.5172 $ & $ -30.4176$ & 10 & Yes\\
                & $3.5353 $ & $ -30.3654$ & 30 & \\
0232.2$-$4420 & $38.1566$ & $ -44.3634$ & 20 & Yes\\
0528.9$-$3927 & $82.2427$ & $ -39.4647$ & 15 & Yes\\
                & $82.3071$ & $ -39.4809$ & 15 & Yes\\
                & $82.1450$ & $ -39.3867$ & 15 & Yes\\
                & $82.1435$ & $ -39.3724$ & 10 & Yes\\
                & $82.1134$ & $ -39.3713$ & 10 & Yes\\
0532.9$-$3701 & $83.1957$ & $ -36.9415$ & 15 & \\
                & $83.1506$ & $ -37.0285$ & 10 & \\
0658.5$-$5556   & $104.5884$& $ -55.9413$& 20 & \\
\hline  
\hline  
  \end{tabular}
  \end{center}
\hspace*{0.3cm}{\footnotesize 
}  } \end{table}

\subsubsection{Comparison with previous results}

Mushotzky \& Scharf (1997) have measured for RXCJ0014.3$-$3022 a
temperature of $12.08^{+1.42}_{-0.88}$~keV (2$\sigma$ errors) with
ASCA data.
% and give an uncertainty boundary of $ \Delta k_{\rm B}T/(k_{\rm B}T) \ltsim 0.3$.
Horner (2001) presents the temperature
for the same cluster of $9.61^{+0.64}_{-0.56}$~keV, 
%(reduced $\chi^2=1.02$ for 1002 d.o.f.),
for RXCJ0232.2$-$4420 of
$7.19^{+0.42}_{-0.38}$keV, 
%(reduced $\chi^2 =1.09$ for 824 d.o.f.), 
for RXCJ0307.0$-$2840 of $6.71^{+0.60}_{-0.53}$keV,
%(reduced $\chi^2 =1.03$ for 491 d.o.f.), 
and for RXCJ1131.9$-$1955 of
$8.26^{+0.63}_{-0.58}$~keV,
%(reduced $\chi^2 =0.94$ for 768 d.o.f.),
all with 2$\sigma$ errors based on ASCA data. Lemonon et al. (1997)
measured a temperature for the post-merger cluster RXCJ1131.9$-$1955
of $5\pm 3$~keV 
%($\chi^2 =18$ for 16 d.o.f.)  
using ROSAT PSPC
data. They also found some evidence for a temperature gradient.

For the cluster RXCJ0658.5$-$5556 the published temperature
measurements are less consistent. Tucker et al. (1998) measured a
temperature of $17.4\pm 2.5$~keV (2$\sigma$ error
%, $\chi^2 =692$ for 778 spectral bins
) with ASCA. Andreani et al. (1999) obtained
$14.5^{+2.0}_{-1.7} $~keV using both ASCA and ROSAT data.  Yaqoob
(1999) measured $11\sim 12$~keV with ASCA data. He found that a
temperature of $\sim 17$~keV can be artificially obtained if the true
spectrum has a stronger low-energy cut-off than that for Galactic
absorption only.

We notice that the ASCA spectra of RXCJ0658.5$-$5556 have in the
0.5--1~keV band only a few data points with large error bars. The
differences in the temperature measurements described above come
therefore clearly from the inclusion or exclusion of this soft part of
the spectrum. Studies of nearby clusters suggest that putative
non-thermal and warm thermal components are important at softer
energies, while for rich clusters like Coma, the ICM dominates the
X-ray emission up to 25~keV (Fusco-Femiano et al. 1999). Non-thermal
emission dominates the emission at energies above 3 keV only in some
of the groups of galaxies (Fukazawa et al. 2000). Since the expected
temperatures of rich clusters are higher than 4~keV, we consider the
temperature determination in the hard energy band as a more robust
measure of the dominant gas mass component of a cluster, which
traces the total mass. For RXCJ0658.5$-$5556 we decided to restrict
our temperature determinations to the energy range
2--12~keV (see Tables~\ref{t:cl_all} and \ref{t:temprofile}).

Our results on RXCJ0658.5$-$5556 are consistent with Chandra
measurements obtained by Markevitch et al. (2002) yielding a
temperature of $14.8^{+1.7}_{-1.2}$~keV from a fit of a spectrum
extracted from the central $ r< 3 ^\prime$ region.  Although the
authors show that $k_{\rm B}T>15$~keV in some parts of this cluster,
the temperatures have quite large error bars of 7~keV so that we regard
this finding as not very significant.

Additionally, Markevitch et al. (2002) fixed the value of the galactic
hydrogen column density to $n_{\rm H}=4.6 \times 10^{20} \; {\rm
cm}^{-2}$ which is significantly lower than $n_{\rm H}=6.5 \times
10^{20} \; {\rm cm}^{-2}$ obtained from Dickey \& Lockman (1999). To
check this result with our XMM-Newton data we set all parameters to be
free to fit the spectrum extracted from the annulus region covering
radii of $0.5<r<4^\prime$ 
in the 0.4--12~keV band. In this case a high temperature
of $k_{\rm B}T=18.8 \pm 2.1$~keV was obtained while the $n_{\rm H}$
went down to an unrealistically small value of $1.8
\pm 0.5\times 10^{20} ~ {\rm cm}^{-2}$, although the remaining
parameters were still relatively reasonable. Therefore, we decide to
fix $n_{\rm H}=6.5 \times 10^{20} ~ {\rm cm}^{-2}$, but exclude the
soft band (0.4--2~keV). In this case we obtained stable results for
the temperature, metallicity, and redshift (see Table~\ref{t:cl_all}).
No significant metallicity gradients were found in our analysis.

\goodbreak
\begin{figure*}
\begin{minipage}{18cm}
\begin{center}
\includegraphics[width=5.9cm]{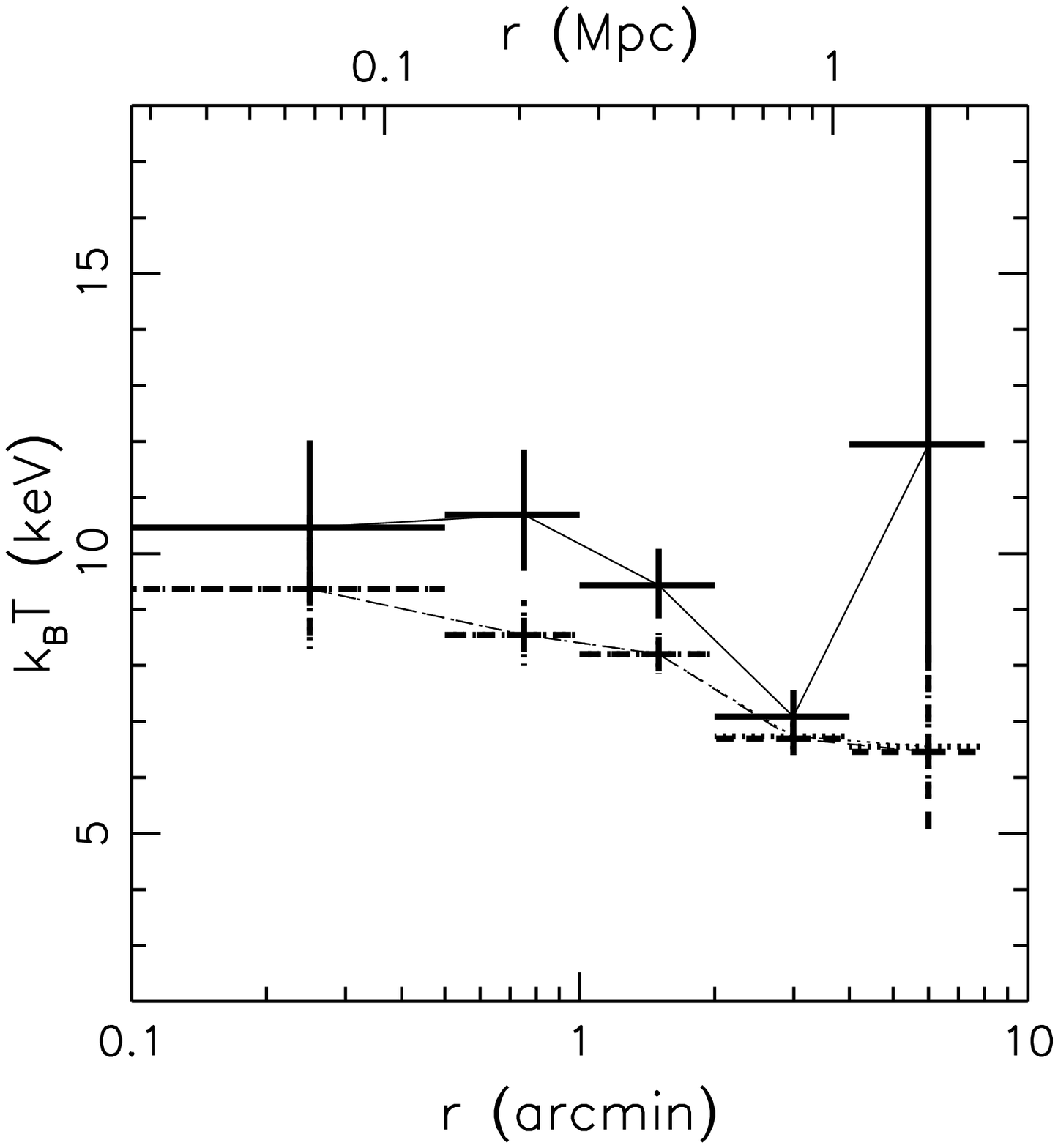}
\includegraphics[width=5.9cm]{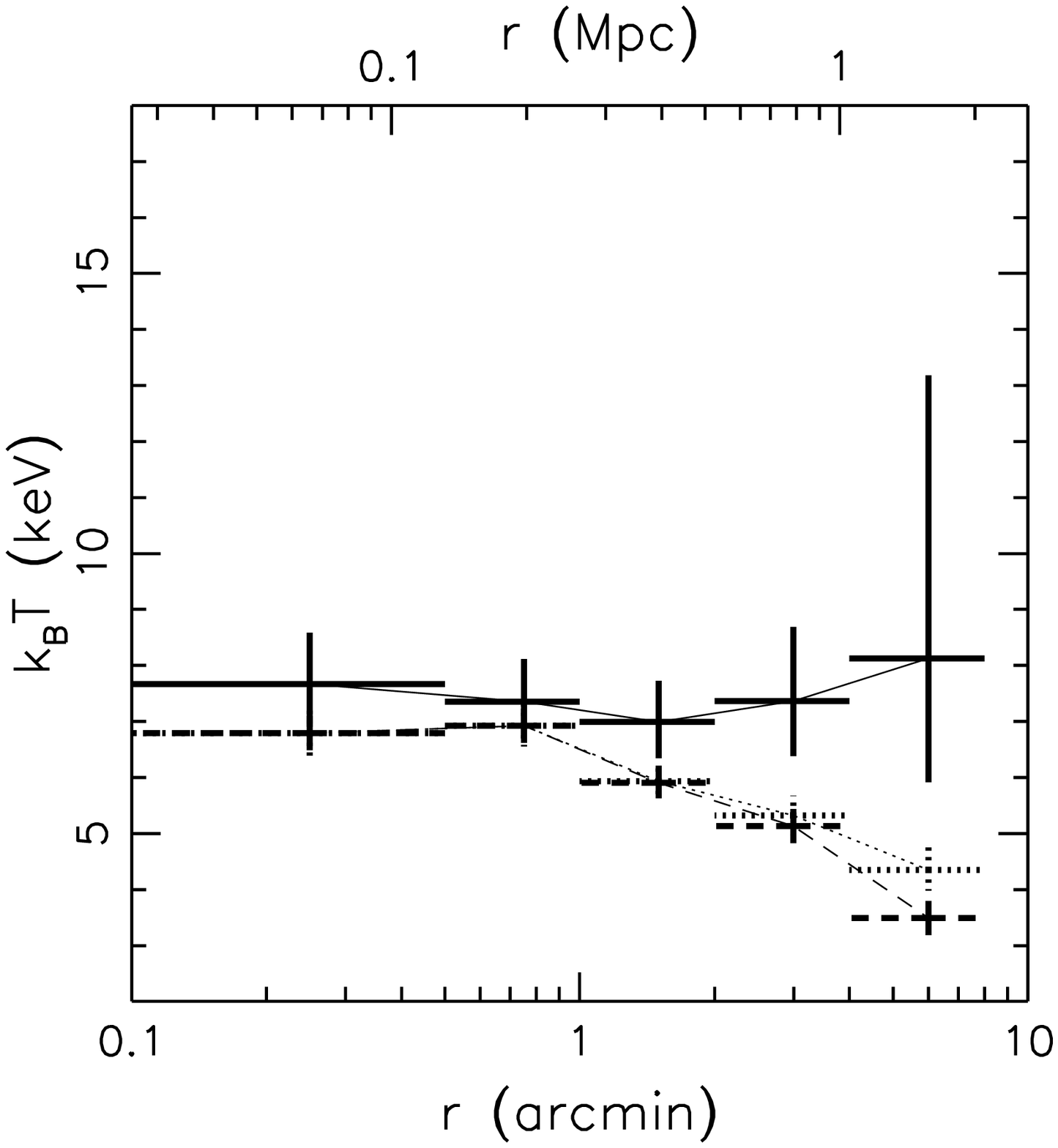}
\includegraphics[width=5.9cm]{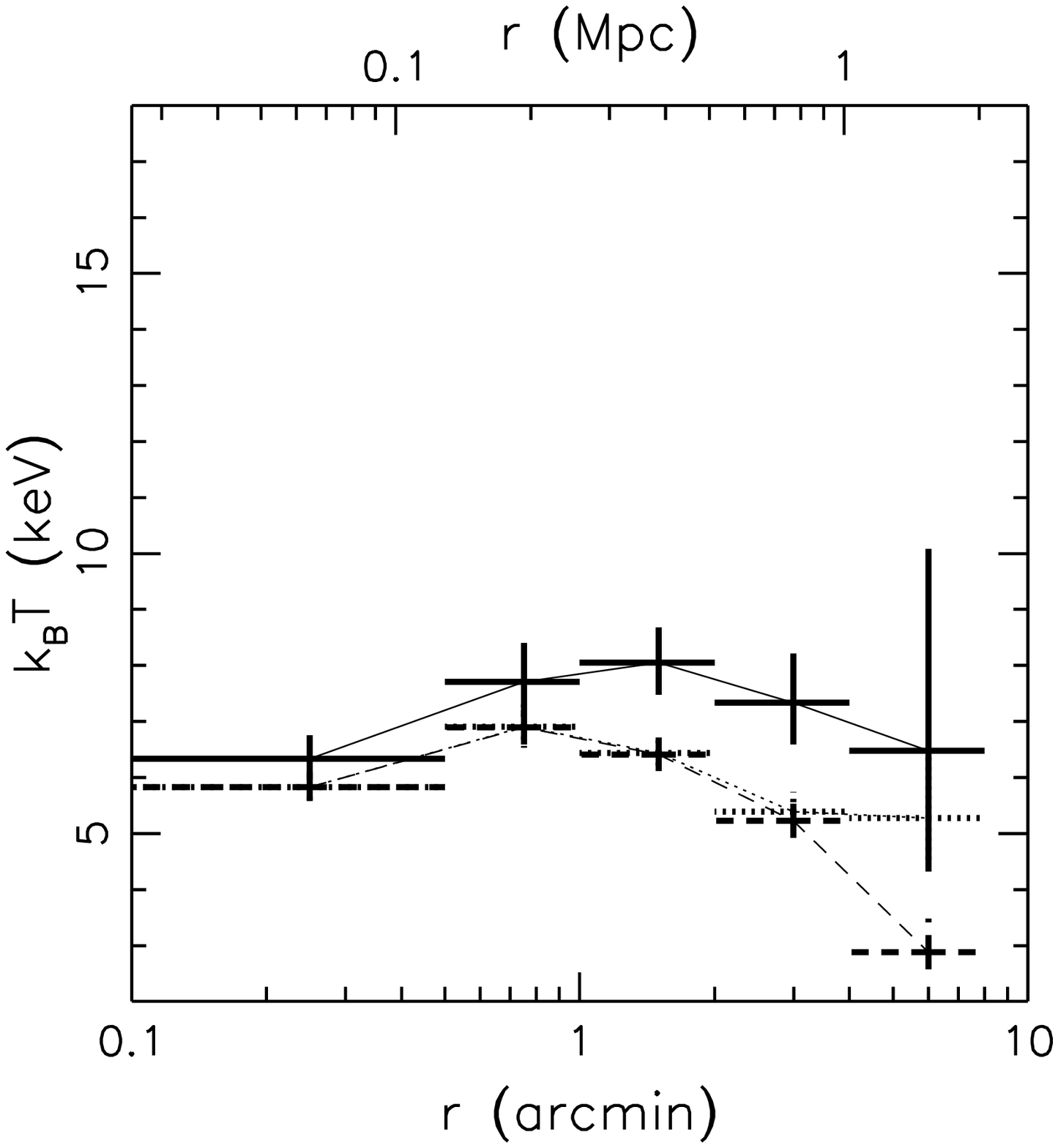}

\includegraphics[width=5.9cm]{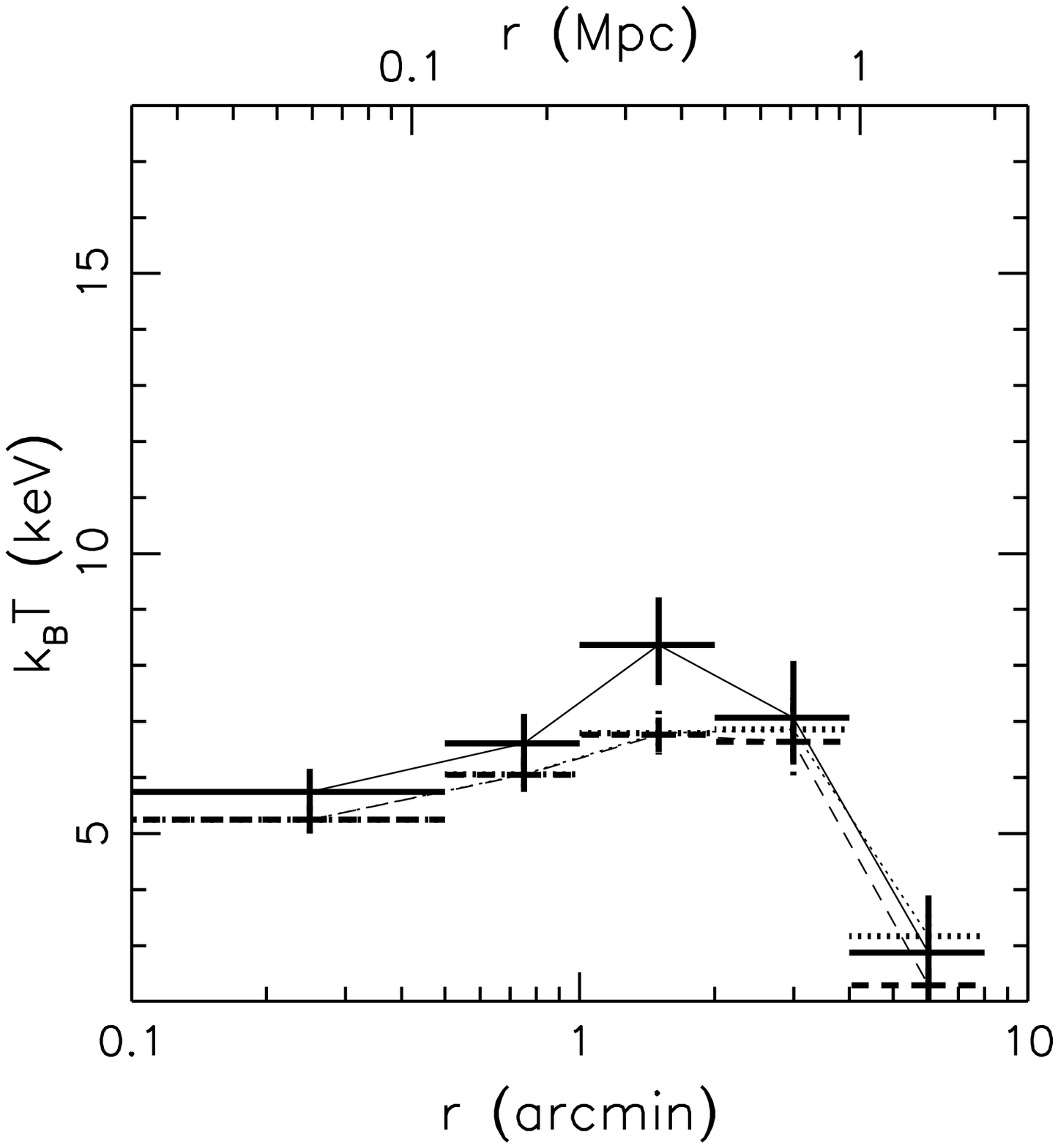}
\includegraphics[width=5.9cm]{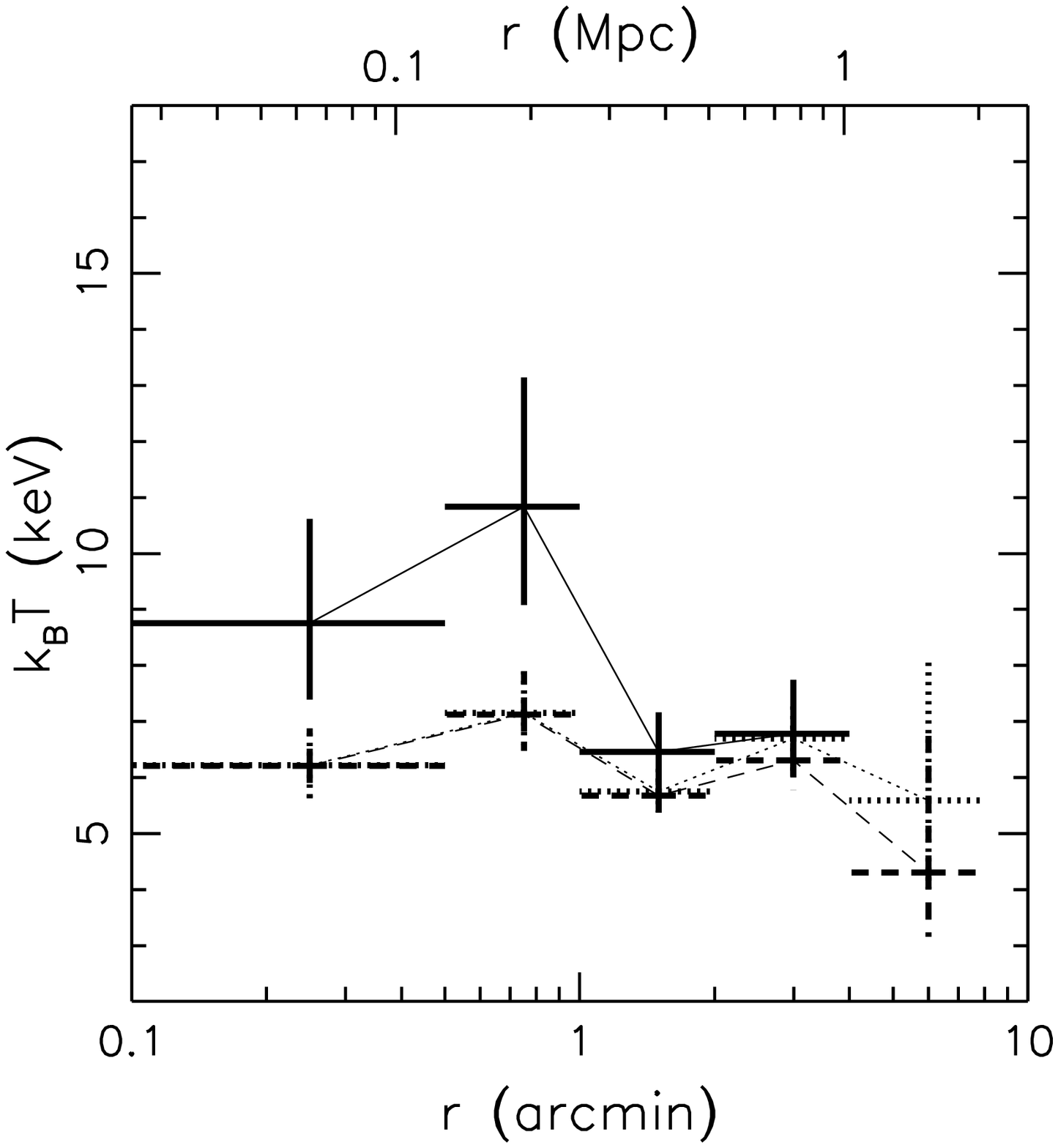}
\includegraphics[width=5.9cm]{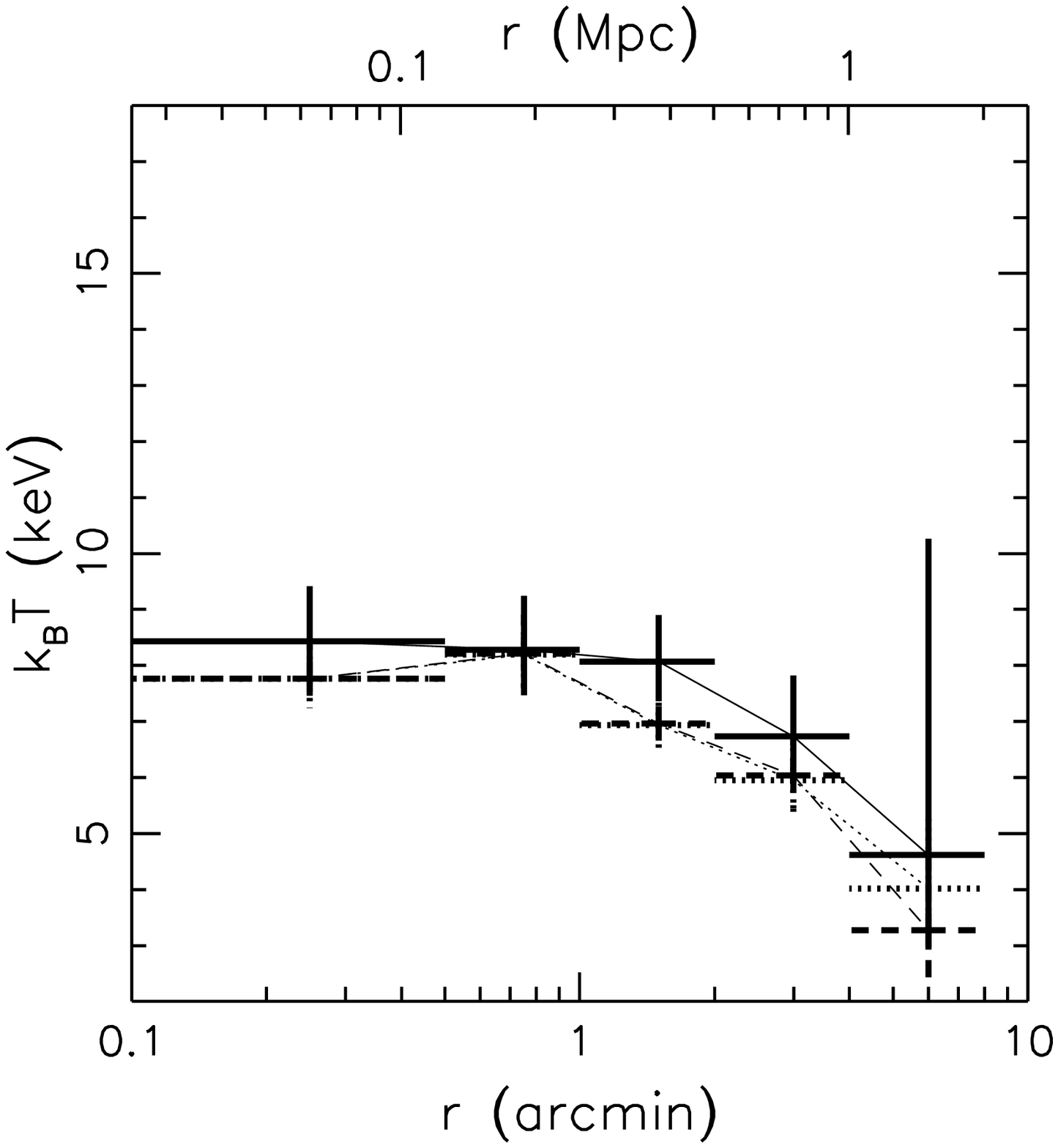}

\includegraphics[width=5.9cm]{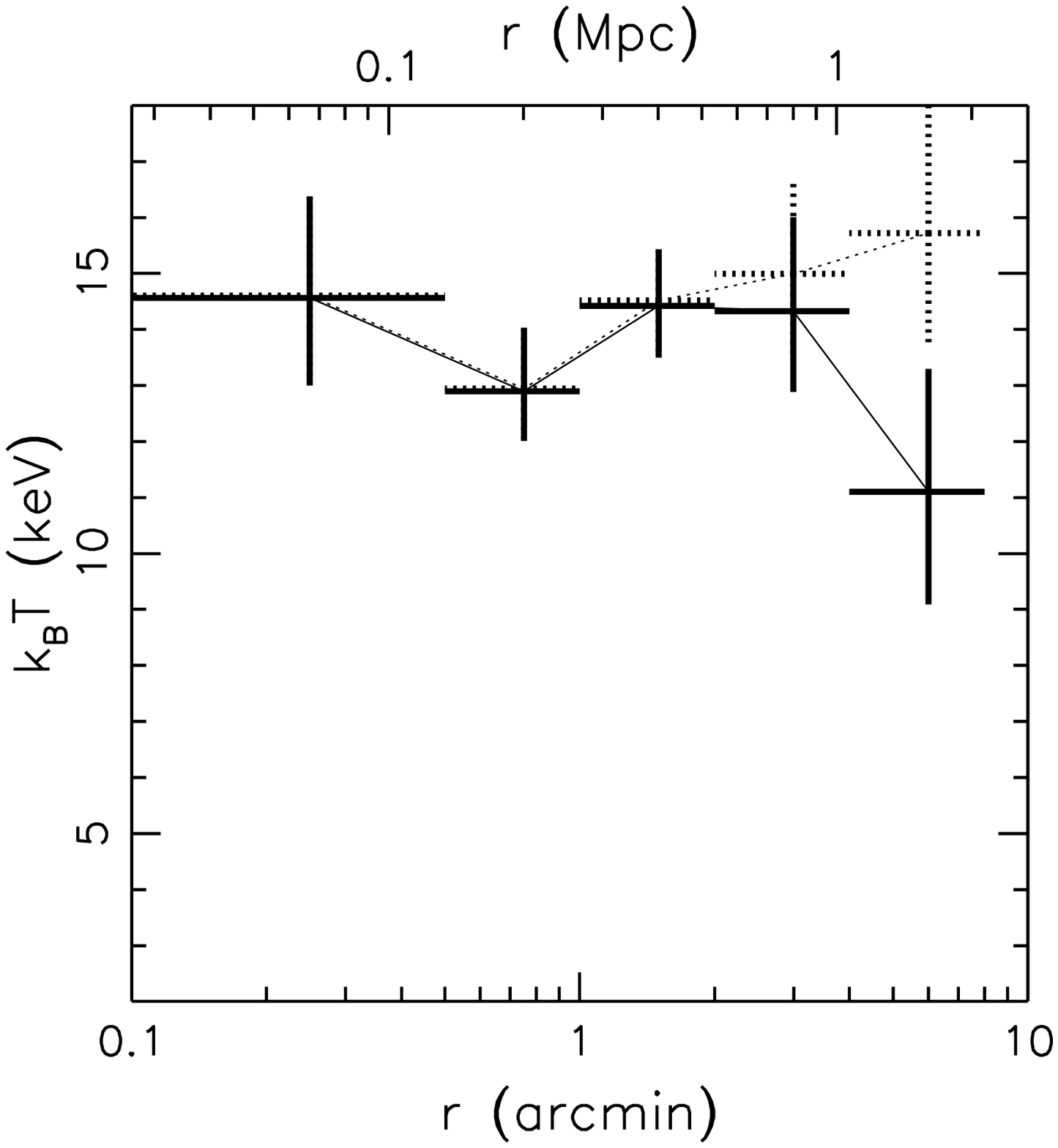}
\includegraphics[width=5.9cm]{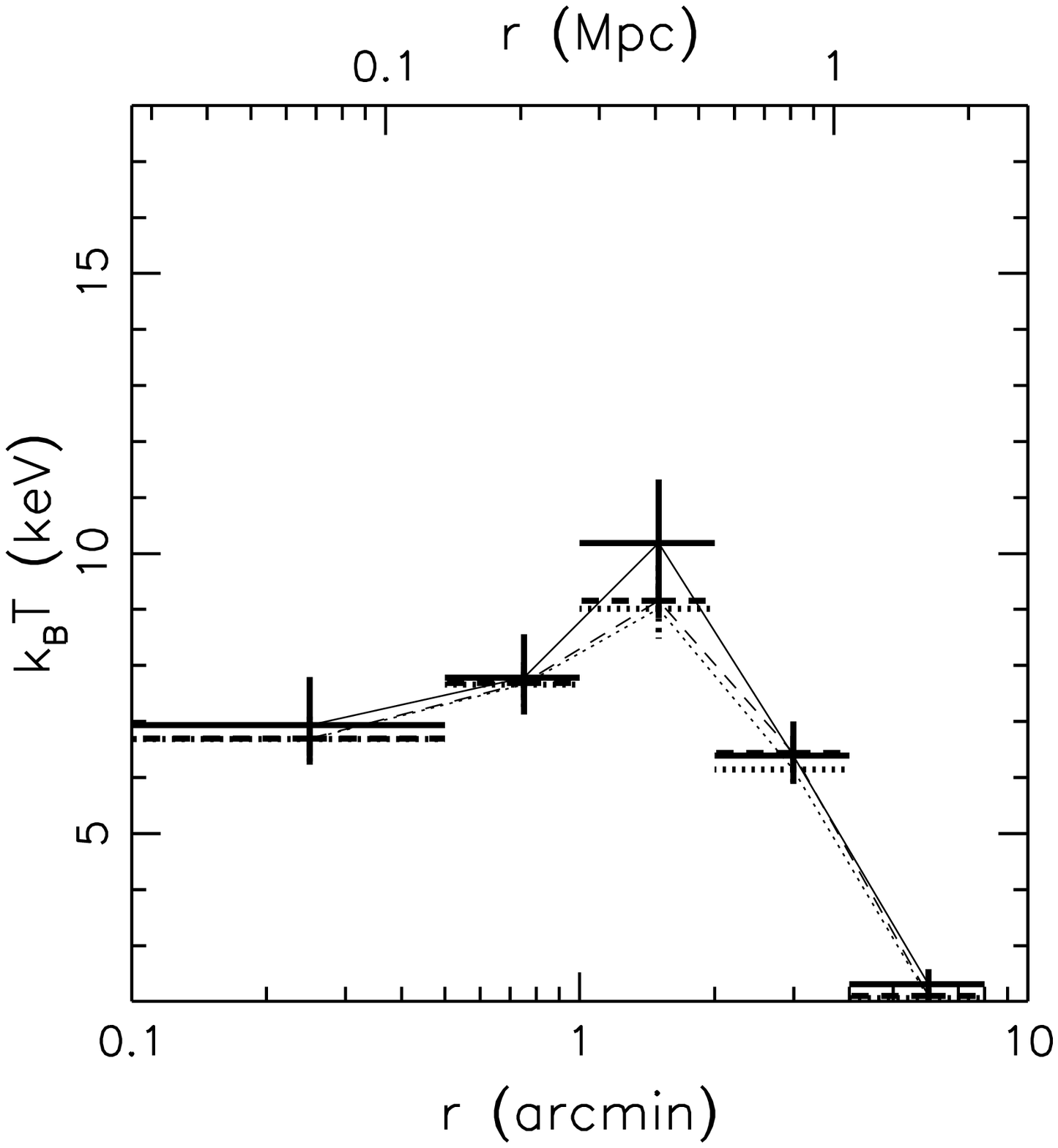}
\includegraphics[width=5.9cm]{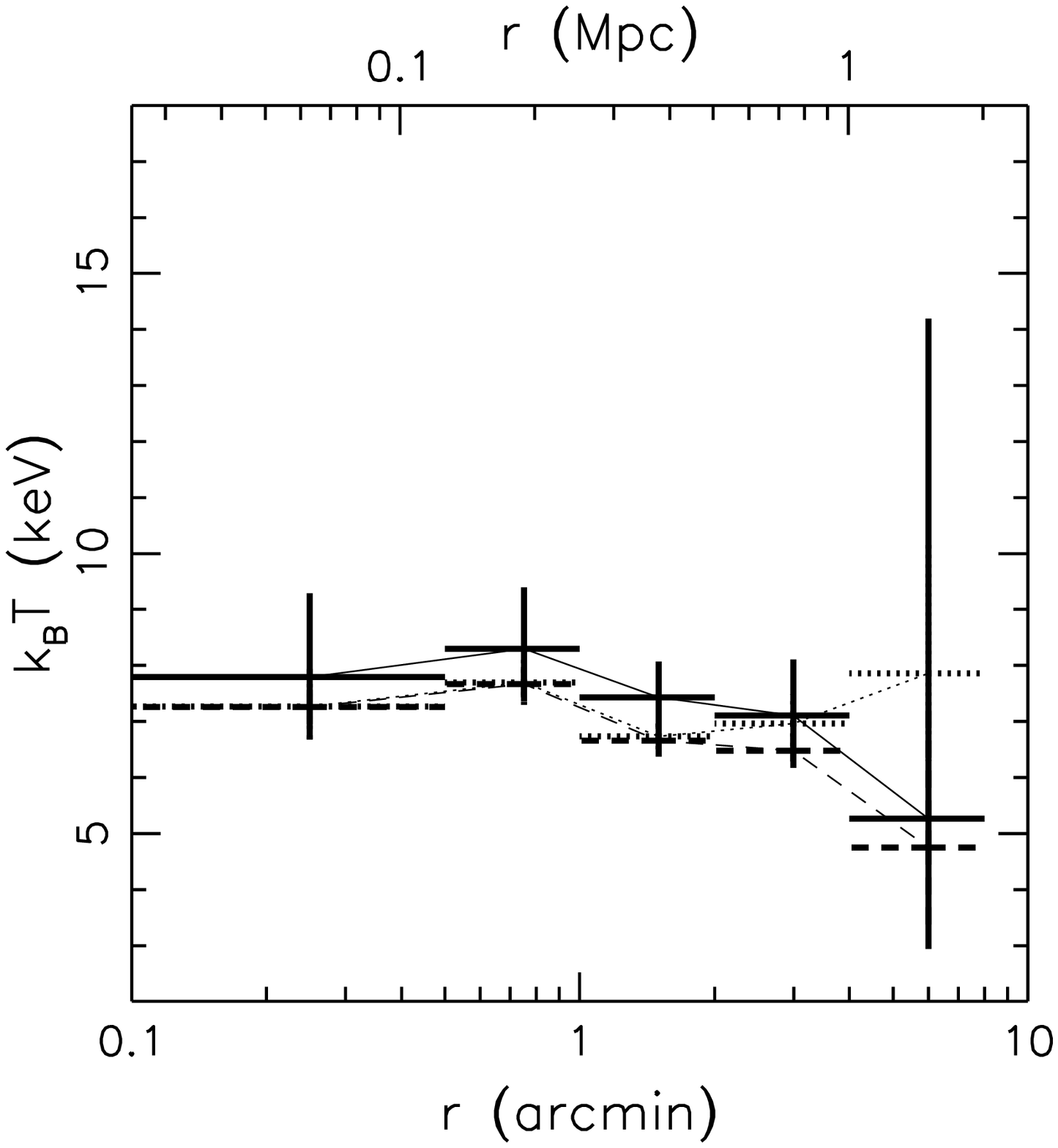}
\end{center}
\end{minipage}
\figcaption{\large From left to right and from top to bottom,
temperature profiles of
RXCJ0014.3$-$3022, RXCJ0043.4$-$2037, 
RXCJ0232.2$-$4420, RXCJ0307.0$-$2840, RXCJ0528.9$-$3927,
RXCJ0532.9$-$3701, RXCJ0658.5$-$5556, RXCJ1131.9$-$1955 
and RXCJ2337.6$+$0016. All the clusters 
except RXCJ0658.5$-$5556 are fitted
in the 0.4--10~keV band with (dashed lines) and without (dotted lines)
the residual background subtraction.  
And the solid lines present the case with the 
residual background subtraction but fitted in the 1--10~keV band.
The corresponding lines show the temperature profiles. 
Temperature profiles of RXCJ0658.5$-$5556 are fitted in 2--12~keV
with (solid lines) and without (dotted lines)
the residual background subtraction. 
\label{f:ktcomp2}
}
\end{figure*}
\goodbreak

\goodbreak
  \begin{table*}
  {
  \begin{center}
  \footnotesize
  {\renewcommand{\arraystretch}{1.3}
  \caption[]{Global temperatures, metallicities and redshifts of REFLEX-DXL
clusters. Col. (1): Cluster name. Col. (2): Radius of annulus in
arcmin. Col. (3): Energy band for fit. Col. (4): X-ray temperature
measurements. Col. (5): Metallicity in solar
abundance. Col. (6): Redshift obtained from the X-ray spectrum. Col. (7):
$\chi^2$ per degree of freedom (d.o.f.). Col. (8): Redshift obtained
from optical spectra as given in the REFLEX catalogue (see B\"ohringer
et al. 2003). All X-ray spectra are fitted by the XSPEC model
``mekal$*$wabs$+$powerlaw/b''.}
  \label{t:cl_all}}
   \begin{tabular}{lllcccrc}
\hline   
\hline     
Cluster          &   Region & Energy band (keV)  & $k_{\rm B}T$ (keV) & Z ($Z_{\odot}$) 
&  $z_{\rm X-ray}$ 
& $ \chi^2/{d.o.f.}$ &  $z_{\rm opt}$ \\ 
\hline 

%redshift is from readkTZz_04781010.f
  RXCJ0014.3-3022 & $0<r<8^\prime$ & $1-10$ &  $   8.65^{+0.43}_{-0.29} $ &  $   0.24 \pm 0.05 $ &  $ 0.294 \pm 0.008 $ &  $   392.9/397 $& $0.3066$ \\
  & $0.5<r<4^\prime$ & $0.4-10$ & $   7.51^{+0.20}_{-0.20} $ &  $   0.22 \pm 0.04 $ &  $ 0.276 \pm 0.009 $ &  $   383.2/371 $ & \\
  $^{*}$ &  &  & $   7.63^{+0.21}_                {-0.20} $ &  $   0.22 \pm 0.04 $ &  $ 0.274 \pm 0.007 $ &  $   377.2/379 $ & \\
  &  & $1-10$ & $   8.29^{+0.43}_                 {-0.32} $ &  $   0.22 \pm 0.05 $ &  $ 0.276 \pm 0.011 $ &  $   269.4/258 $ & \\
 RXCJ0043.4-2037 & $0<r<8^\prime$ & $1-10$ &  $   7.50^{+0.47}_{-0.40} $ &  $   0.23 \pm 0.06 $ &  $ 0.309 \pm 0.015 $ &  $   198.0/238 $& $0.2924$ \\
  & $0.5<r<4^\prime$ & $0.4-10$ & $   5.88^{+0.22}_{-0.21} $ &  $   0.24 \pm 0.06 $ &  $ 0.303\pm0.006 $ &  $   197.8/197 $ & \\
  $^{*}$ &  &  & $   5.96^{+0.23}_                {-0.21} $ &  $   0.25 \pm 0.06 $ &  $ 0.302 \pm 0.011 $ &  $   191.1/197 $ & \\
  &  & $1-10$ & $   6.81^{+0.43}_                 {-0.39} $ &  $   0.23 \pm 0.07 $ &  $ 0.300 \pm 0.012 $ &  $   124.4/129 $ & \\
 RXCJ0232.2-4420 & $0<r<8^\prime$ & $1-10$ &  $   6.70^{+0.24}_{-0.25} $ &  $   0.32 \pm 0.05 $ &  $ 0.275 \pm 0.002 $ &  $   292.0/332 $& $0.2836$ \\
  & $0.5<r<4^\prime$ & $0.4-10$ & $   6.33^{+0.22}_{-0.17} $ &  $   0.22 \pm 0.05 $ &  $ 0.295 \pm 0.008 $ &  $   384.2/287 $ & \\
  $^{*}$ &  &  & $   6.27^{+0.20}_                {-0.17} $ &  $   0.23 \pm 0.05 $ &  $ 0.296 \pm 0.010 $ &  $   372.6/287 $ & \\
  &  & $1-10$ & $   7.62^{+0.40}_                 {-0.33} $ &  $   0.24 \pm 0.05 $ &  $ 0.296 \pm 0.009 $ &  $   195.4/195 $ & \\
 RXCJ0307.0-2840 & $0<r<8^\prime$ & $1-10$ &  $   6.17^{+0.24}_{-0.27} $ &  $   0.28 \pm 0.06 $ &  $ 0.241 \pm 0.002 $ &  $   270.8/281 $& $0.2578$ \\
  & $0.5<r<4^\prime$ & $0.4-10$ & $   6.10^{+0.20}_{-0.22} $ &  $   0.28 \pm 0.06 $ &  $ 0.244 \pm 0.003 $ &  $   275.0/251 $ & \\
  $^{*}$ &  &  & $   6.17^{+0.21}_                {-0.21} $ &  $   0.28 \pm 0.06 $ &  $ 0.244 \pm 0.004 $ &  $   265.1/251 $ & \\
  &  & $1-10$ & $   6.63^{+0.34}_                 {-0.31} $ &  $   0.29 \pm 0.07 $ &  $ 0.243 \pm 0.004 $ &  $   176.5/167 $ & \\
 RXCJ0528.9-3927 & $0<r<8^\prime$ & $1-10$ &  $   8.07^{+0.59}_{-0.52} $ &  $   0.31 \pm 0.08 $ &  $ 0.262 \pm 0.012 $ &  $   163.2/159 $& $0.2839$ \\
  & $0.5<r<4^\prime$ & $0.4-10$ & $   6.49^{+0.29}_{-0.27} $ &  $   0.30 \pm 0.08 $ &  $ 0.281 \pm 0.011 $ &  $   162.8/148 $ & \\
  $^{*}$ &  &  & $   6.42^{+0.28}_                {-0.27} $ &  $   0.31 \pm 0.08 $ &  $ 0.281 \pm 0.010 $ &  $   154.8/148 $ & \\
  &  & $1-10$ & $   7.66^{+0.63}_                 {-0.52} $ &  $   0.28 \pm 0.10 $ &  $ 0.282 \pm 0.016 $ &  $    91.5/ 97 $ & \\
 RXCJ0532.9-3701 & $0<r<8^\prime$ & $1-10$ &  $   7.46^{+0.38}_{-0.36} $ &  $   0.34 \pm 0.07  $ &  $ 0.274 \pm 0.003 $ &  $   286.1/239 $& $0.2747$ \\
  & $0.5<r<4^\prime$ & $0.4-10$ & $   7.09^{+0.31}_{-0.44} $ &  $   0.29 \pm 0.08 $ &  $ 0.262 \pm 0.009 $ &  $   226.5/202 $ & \\
  $^{*}$ &  &  & $   6.98^{+0.35}_                {-0.28} $ &  $   0.30 \pm 0.09 $ &  $ 0.262 \pm 0.011 $ &  $   224.3/202 $ & \\
  &  & $1-10$ & $   7.76^{+0.72}_                 {-0.48} $ &  $   0.28 \pm 0.10  $ &  $ 0.259 \pm 0.016 $ &  $   143.7/137 $ & \\
 RXCJ0658-5557   & $0<r<8^\prime$ & $2-12$ &  $  13.59^{+0.71}_{-0.58} $ &  $   0.24 \pm 0.03 $ &  $ 0.287 \pm 0.002 $ &  $   621.3/608 $& $0.2965$ \\
  & $0.5<r<4^\prime$ & $2-12$ & $  14.56^{+0.91}_{-0.69} $ &  $   0.21 \pm 0.04 $ &  $ 0.290 \pm 0.005 $ &  $   378.1/400 $ & \\
  $^{*}$ &  &  & $  14.61^{+0.57}_                {-0.74} $ &  $   0.21 \pm 0.04 $ &  $ 0.291 \pm 0.004 $ &  $   377.1/400 $ & \\
  $^{*}$ &  & 0.4-10 & $   9.63^{+0.22}_          {-0.17} $ &  $   0.23 \pm 0.02 $ &  $ 0.288 \pm 0.002 $ &  $  1214.5/892 $ & \\
 RXCJ1131.9-1955 & $0<r<8^\prime$ & $1-10$ &  $   6.42^{+0.26}_{-0.25} $ &  $   0.26 \pm 0.05 $ &  $ 0.299 \pm 0.003 $ &  $   322.5/283 $& $0.3075$ \\
  & $0.5<r<4^\prime$ & $0.4-10$ & $   6.71^{+0.08}_{-0.40} $ &  $   0.23 \pm 0.09 $ &  $ 0.285 \pm 0.009 $ &  $   263.1/255 $ & \\
  $^{*}$ &  &  & $   6.62^{+0.38}_                {-0.21} $ &  $   0.26 \pm 0.07 $ &  $ 0.287 \pm 0.012 $ &  $   248.4/255 $ & \\
  &  & $1-10$ & $   7.44^{+0.67}_                 {-0.22} $ &  $   0.26 \pm 0.07 $ &  $ 0.295 \pm 0.024 $ &  $   163.4/174 $ & \\
 RXCJ2337.6+0016 & $0<r<8^\prime$ & $1-10$ &  $   8.02^{+0.59}_{-0.57} $ &  $   0.22 \pm 0.07 $ &  $ 0.328 \pm 0.009 $ &  $   211.1/255 $& $0.2779$ \\
  & $0.5<r<4^\prime$ & $0.4-10$ & $   6.77^{+0.29}_{-0.24} $ &  $   0.24 \pm 0.06 $ &  $ 0.315 \pm 0.011 $ &  $   176.5/234 $ & \\
  $^{*}$ &  &  & $   6.83^{+0.22}_                {-0.28} $ &  $   0.25 \pm 0.06 $ &  $ 0.318 \pm 0.009 $ &  $   172.7/234 $ & \\
  &  & $1-10$ & $   7.50^{+0.44}_                 {-0.37} $ &  $   0.22 \pm 0.06 $ &  $ 0.317 \pm 0.011 $ &  $   116.2/161 $ & \\

\hline  
\hline  
  \end{tabular}
  \end{center}
\hspace*{0.3cm}{\footnotesize }\\
$^{*}$Introduce the upper limit of the residual Galactic emission described by 
an ``apec'' model in the double background subtraction
method.
}
\end{table*}
\goodbreak

\begin{figure}
\begin{center}
\includegraphics[width=8.5cm]
{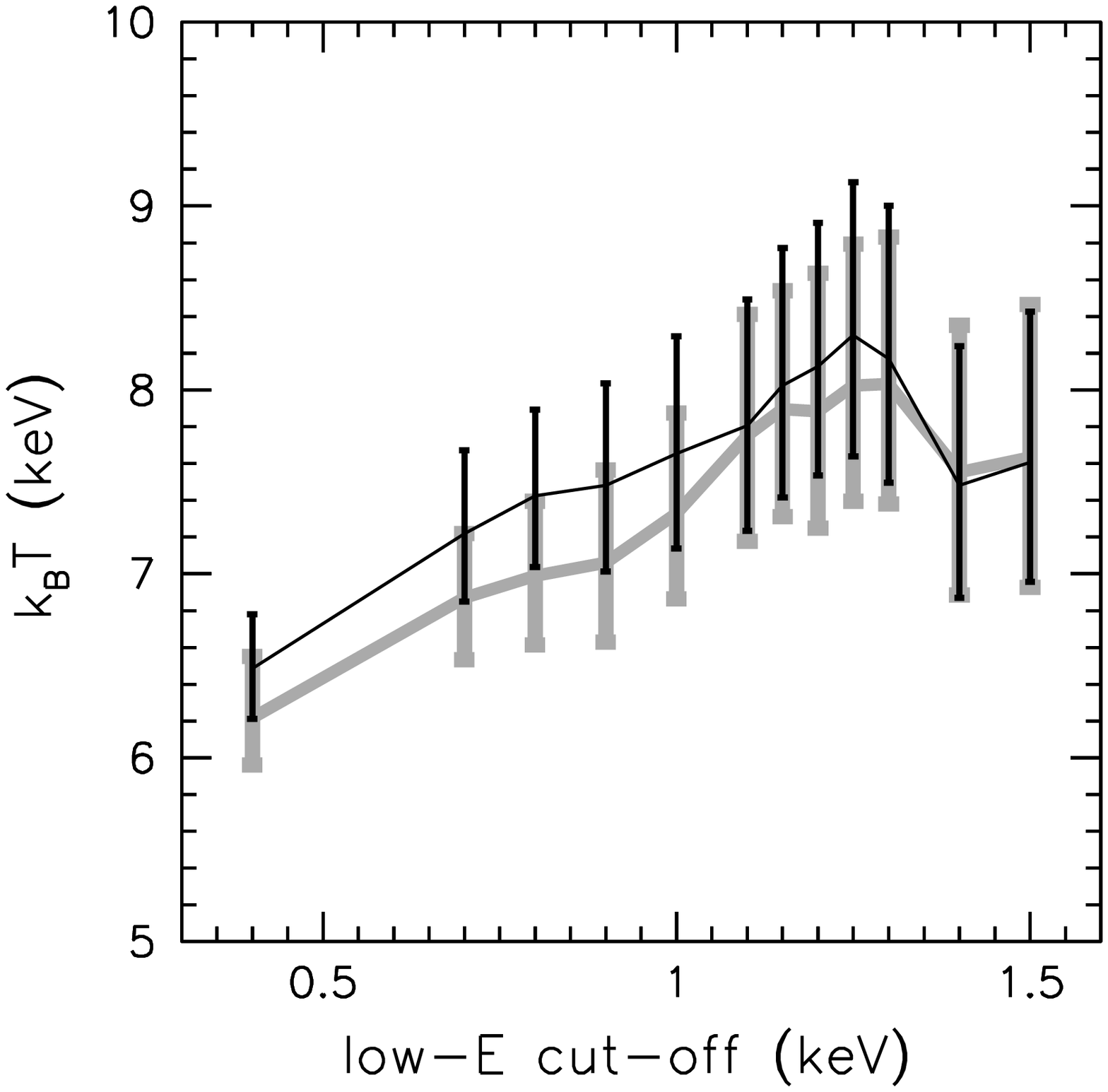}
\end{center}
\figcaption{The temperature 
measurements versus low-E cut-off from two different methods, 
the double background subtraction method (black)
and the method applied in Arnaud et al. 2002 (grey).
\label{f:ktdeviation}
}
\end{figure}

We notice that the temperature of RXCJ0528.9$-$3927
(also in other clusters) changes 
significantly with the low cut-off of the energy band used in the
fit. In Fig.~\ref{f:ktdeviation},
we thus used the X-ray spectra to test the energy band dependence
and possible method dependencies by comparing the temperature 
measurements versus low energy band (low-E) cut-off from two different methods 
(the double background subtraction method
and the method applied in Arnaud et al. 2002, i.e. 
we use the standard XMMSAS command `evigweight' 
to correct vignetting, 
use `arfgen' and `rmfgen' to create on-axis arf and rmf, 
and apply the blank sky provided by Lumb 2002). For
comparison, we applied the following models to fit the spectra in 
the 0.4--10~keV band after double background subtraction. 
We obtain $k_{\rm B}T=7.69^{+0.28}_{-0.46}$~keV and 
$n_{\rm H} \sim 0-0.1 \times 10^{20}~{\rm cm^{-2}}$ using 
a single-phase temperature model (``wabs$*$mekal$+$powerlaw/b'') 
with free $n_{\rm H}$. We obtain 
$k_{\rm B}T=10.12 \pm 3.13$~keV using two component thermal model 
(``wabs$*$(mekal$+$powerlaw)$+$powerlaw/b'') with fixed $n_{\rm H}$.
We also obtain
$k_{\rm B}T_1=9.34^{+5.93}_{-1.04}$~keV and $k_{\rm B}T_2~\sim~0.49-1.34$~keV
using two thermal component model 
(``wabs$*$(mekal$+$apec)$+$powerlaw/b''), in which we fix
the redshift of the soft component to the redshift of the cluster. 
The metallicity and 
redshift measurements among the different modelings and different 
low-E cut-off vary within 5\%. The results presented in
Fig.~\ref{f:ktdeviation} suggest some influence of the low energy
band on the temperature measurements. As we discussed above, the results
obtained in the harder energy band should recover the correct cluster
temperature. Similar phenomena are also found for A1413
(Pratt \& Arnaud 2002) using XMM-Newton data, and Coma, A1795 and
A3112 (Nevalainen et al. 2003) based on the comparison of XMM-Newton
and ROSAT PSPC observations. Nevalainen et al. interprete this as a
soft excess, possibly due to a `warm-hot' intergalactic medium.  We
will analyse this feature of our sample in more detail in a
forthcoming paper.

\subsection{Temperature profiles}
\label{s:tem}

We have already noted that the differences between the global
temperatures of the regions covering radii of 
$0.5<r<4^\prime$ and $r<8^\prime$, respectively, in
Table~\ref{t:cl_all} are possibly caused by systematic temperature
gradients. For a more detailed study of the temperature profiles we
devide the cluster regions into the five annuli 0--0.5$^\prime$,
0.5--1$^\prime$, 1--2$^\prime$, 2--4$^\prime$, and 4--8$^\prime$
(cf. Fig.~\ref{f:ktprof}). Note that in the spectra extracted from the
outermost rings of RXCJ0014.3$-$3022 and RXCJ0528.9$-$3927 we ignore a
narrow energy band containing the residual background around the
1.49~keV Al line (Freyberg et al. 2002).

Table~\ref{t:temprofile} shows the temperature profile catalogue of
the clusters from the spectral analysis fitted by 
one or two step background subtraction (``mekal$*$wabs'',
``mekal$*$wabs$+$powerlaw/b''). We use both the $>0.4$~keV and
$>1$~keV energy bands except for RXCJ0658.5$-$5556. We
apply the 2--12~keV band for this high temperature cluster.  The
temperature profiles of the clusters are also shown in
Fig.~\ref{f:ktcomp2}.

\goodbreak
  \begin{table*}
  {
  \begin{center}
  \footnotesize
  {\renewcommand{\arraystretch}{1.3}
  \caption[]{Temperature profiles of REFLEX-DXL clusters obtained from 
the spectral analysis with residual background subtraction (double
background subtraction) and without residual background
subtraction. Col.(1): Cluster name. Col.(2): Energy band for spectral
fit. Col.(3): Model used in XSPEC. Cols.(4-8): Temperature measurements.
We do not obtain a consistent temperature measurement for RXCJ0528.9$-$3927
in 4--8$^\prime$ region from the combined data 
because of the high background. }
  \label{t:temprofile}}

  \begin{tabular}{llccccccccc}
\hline   
\hline     
Cluster          & Energy band & Model & \multicolumn{5}{c}{$k_{\rm B}T$ (keV)} \\ 
                 & (keV)       &       & $0<r<0.5^\prime$ & $0.5<r<1^\prime$ & 
                 $1<r<2^\prime$ & $2<r<4^\prime$ & $4<r<8^\prime$  \\  
\hline   
 RXCJ0014.3$-$3022 & 0.4-10 & model1$^{a}$ & $  9.37^{+  1.30}_{-  1.07}$ &
 $  8.54^{+  0.62}_{-  0.54}$ &
 $  8.20^{+  0.38}_{-  0.35}$ &
 $  6.74^{+  0.32}_{-  0.29}$ &
 $  6.56^{+  1.30}_{-  1.02}$\\
                 &        & model2$^{b}$ & $  9.37^{+  1.31}_{-  1.07}$ &
 $  8.55^{+  0.62}_{-  0.55}$ &
 $  8.20^{+  0.39}_{-  0.35}$ &
 $  6.70^{+  0.32}_{-  0.30}$ &
 $  6.46^{+  2.14}_{-  1.49}$ \\
                 & 1.0-10 & model2 &$  10.46^{+  1.56}_{-  1.33}$ &
 $  10.69^{+  1.17}_{-  1.00}$ &
 $  9.43^{+  0.65}_{-  0.59}$ &
 $  7.09^{+  0.47}_{-  0.43}$ &
 $  11.94^{+  11.69}_{-  4.03}$ \\
 RXCJ0043.4$-$2037 & 0.4-10 & model1 & $  6.79^{+ 0.47}_{- 0.41} $ & $  6.93^{+ 0.42}_{- 0.37} $ & $  5.93^{+ 0.31}_{- 0.28} $ & $  5.33^{+ 0.36}_{- 0.34} $ & $  4.35^{+ 0.43}_{- 0.37} $ \\
                 &        & model2 & $  6.79^{+ 0.47}_{- 0.42} $ & $  6.93^{+ 0.42}_{- 0.37} $ & $  5.91^{+ 0.31}_{- 0.28} $ & $  5.13^{+ 0.38}_{- 0.34} $ & $  3.49^{+ 0.46}_{- 0.35} $ \\
                 & 1.0-10 & model2 & $  7.67^{+ 0.92}_{- 0.76} $ & $  7.36^{+ 0.76}_{- 0.66} $ & $  6.99^{+ 0.74}_{- 0.65} $ & $  7.37^{+ 1.32}_{- 0.99} $ & $  8.13^{+ 5.05}_{- 2.21} $ \\
 RXCJ0232.2$-$4420 & 0.4-10 & model1 & $  5.83^{+ 0.27}_{- 0.25} $ & $  6.91^{+ 0.42}_{- 0.37} $ & $  6.44^{+ 0.32}_{- 0.29} $ & $  5.39^{+ 0.35}_{- 0.34} $ & $  5.28^{+ 1.19}_{- 0.86} $ \\
                 &        & model2 & $  5.83^{+ 0.27}_{- 0.25} $ & $  6.90^{+ 0.42}_{- 0.37} $ & $  6.41^{+ 0.32}_{- 0.30} $ & $  5.23^{+ 0.36}_{- 0.33} $ & $  2.88^{+ 0.61}_{- 0.37} $ \\                
                 & 1.0-10 & model2 & $  6.34^{+ 0.42}_{- 0.37} $ & $  7.71^{+ 0.69}_{- 0.60} $ & $  8.05^{+ 0.63}_{- 0.57} $ & $  7.34^{+ 0.87}_{- 0.74} $ & $  6.48^{+ 3.61}_{- 2.16} $ \\
 RXCJ0307.0$-$2840 & 0.4-10 & model1 & $  5.25^{+  0.26}_{-  0.24}$ &
 $  6.06^{+  0.34}_{-  0.31}$ &
 $  6.80^{+  0.42}_{-  0.39}$ &
 $  6.86^{+  0.69}_{-  0.61}$ &
 $  3.17^{+  0.50}_{-  0.40}$\\
                 &        & model2 & $  5.24^{+  0.26}_{-  0.24}$ & $  6.05^{+  0.34}_{-  0.31}$ & $  6.76^{+  0.42}_{-  0.39}$ &
 $  6.64^{+  0.70}_{-  0.60}$ &
 $  2.30^{+  0.40}_{-  0.30}$ \\
                 & 1.0-10 & model2 &  $  5.74^{+  0.40}_{-  0.36}$ &
 $  6.61^{+  0.53}_{-  0.46}$ &
 $  8.37^{+  0.84}_{-  0.72}$ &
 $  7.07^{+  1.01}_{-  0.84}$ &
 $  2.87^{+  1.02}_{-  0.63}$ \\
 RXCJ0528.9$-$3927 & 0.4-10 & model1 & $  6.22^{+ 0.67}_{- 0.58} $ & $  7.16^{+ 0.77}_{- 0.65} $ & $  5.75^{+ 0.43}_{- 0.38} $ & $  6.70^{+ 0.68}_{- 0.58} $ & $  5.59^{+ 2.56}_{- 1.58} $ \\
                 &        & model2 & $  6.21^{+ 0.67}_{- 0.58} $ & $  7.12^{+ 0.76}_{- 0.64} $ & $  5.68^{+ 0.42}_{- 0.37} $ & $  6.30^{+ 0.62}_{- 0.53} $ & $  4.30^{+ 2.67}_{- 1.14} $ \\
                 & 1.0-10 & model2 & $  8.75^{+ 1.86}_{- 1.36} $ & $ 10.84^{+ 2.31}_{- 1.76} $ & $  6.46^{+ 0.70}_{- 0.59} $ & $  6.79^{+ 0.96}_{- 0.77} $ &  \\
 RXCJ0532.9$-$3701 & 0.4-10 & model1 & $  7.77^{+ 0.61}_{- 0.53} $ & $  8.19^{+ 0.70}_{- 0.61} $ & $  6.93^{+ 0.48}_{- 0.43} $ & $  5.96^{+ 0.64}_{- 0.56} $ & $  4.02^{+ 1.24}_{- 0.81} $ \\
                 &        & model2 & $  7.77^{+ 0.61}_{- 0.53} $ & $  8.21^{+ 0.71}_{- 0.61} $ & $  6.96^{+ 0.49}_{- 0.43} $ & $  6.04^{+ 0.69}_{- 0.59} $ & $  3.28^{+ 1.34}_{- 0.89} $ \\
                 & 1.0-10 & model2 & $  8.44^{+ 0.98}_{- 0.81} $ & $  8.28^{+ 0.97}_{- 0.80} $ & $  8.07^{+ 0.83}_{- 0.71} $ & $  6.73^{+ 1.09}_{- 0.85} $ & $  4.61^{+ 5.65}_{- 1.69} $ \\
 RXCJ0658.5$-$5556   & 2.0-12 & model1 & $  14.60^{+ 1.81}_{- 1.57} $ & $  12.94^{+ 1.13}_{- 0.90} $ & $  14.52^{+ 0.98}_{- 0.91} $ & $  14.99^{+ 1.62}_{- 1.46} $ & $  15.72^{+ 2.40}_{- 2.01} $ \\
                 & 2.0-12 &model2 & $  14.57^{+ 1.57}_{- 1.80} $ & $  12.90^{+ 1.11}_{- 0.89} $ & $  14.42^{+ 0.98}_{- 0.92} $ & $  14.33^{+ 1.68}_{- 1.44} $ & $  11.1^{+ 2.20}_{- 2.00} $ \\
 RXCJ1131.9$-$1955 & 0.4-10 & model1 & $  6.69^{+ 0.46}_{- 0.41} $ & $  7.66^{+ 0.47}_{- 0.42} $ & $  9.02^{+ 0.61}_{- 0.54} $ & $  6.15^{+ 0.36}_{- 0.33} $ & $  2.06^{+ 0.11}_{- 0.10} $ \\
                 &        & model2 & $  6.70^{+ 0.47}_{- 0.41} $ & $  7.69^{+ 0.47}_{- 0.42} $ & $  9.15^{+ 0.63}_{- 0.56} $ & $  6.44^{+ 0.41}_{- 0.36} $ & $  2.10^{+ 0.14}_{- 0.12} $ \\ 
                 & 1.0-10 & model2 & $  6.93^{+ 0.86}_{- 0.70} $ & $  7.79^{+ 0.77}_{- 0.66} $ & $ 10.19^{+ 1.14}_{- 0.94} $ & $  6.39^{+ 0.61}_{- 0.51} $ & $  2.31^{+ 0.27}_{- 0.22} $ \\
 RXCJ2337.6$+$0016 & 0.4-10 & model1 & $  7.27^{+ 0.67}_{- 0.58} $ & $  7.70^{+ 0.48}_{- 0.43} $ & $  6.73^{+ 0.31}_{- 0.29} $ & $  6.96^{+ 0.53}_{- 0.46} $ & $  7.86^{+ 2.34}_{- 1.56} $ \\
                 &        & model2 & $  7.26^{+ 0.68}_{- 0.58} $ & $  7.67^{+ 0.48}_{- 0.43} $ & $  6.66^{+ 0.31}_{- 0.29} $ & $  6.48^{+ 0.49}_{- 0.44} $ & $  4.75^{+ 2.52}_{- 1.37} $ \\
                 & 1.0-10 & model2 & $  7.80^{+ 1.49}_{- 1.11} $ & $  8.30^{+ 1.09}_{- 0.86} $ & $  7.43^{+ 0.64}_{- 0.55} $ & $  7.11^{+ 1.00}_{- 0.82} $ & $  5.26^{+ 8.93}_{- 2.33} $ \\
\hline  
\hline  
  \end{tabular}
  \end{center}
$^{a}$ \hspace*{0.3cm}{\footnotesize mekal$*$wabs.}\\
$^{b}$ \hspace*{0.3cm}{\footnotesize mekal$*$wabs$+$powerlaw/b.}\\
}
  \end{table*}
\goodbreak

For most of the objects, the temperature can be measured out to
$r_{500}$.  The overall temperature profile is characterized by a
rather moderate decrease towards the center, and a decrease towards
the outer regions, yet on different levels, including no decrease at
all for some of the clusters. This confirms a suggestion of Finoguenov
et al.  (2001b) that the differences in the behaviour of the
temperature profiles in the outskirts of clusters have a statistical
origin, rather than simple reflections of measurement errors.

\begin{figure}
\begin{center}
\includegraphics[width=8.4cm]
{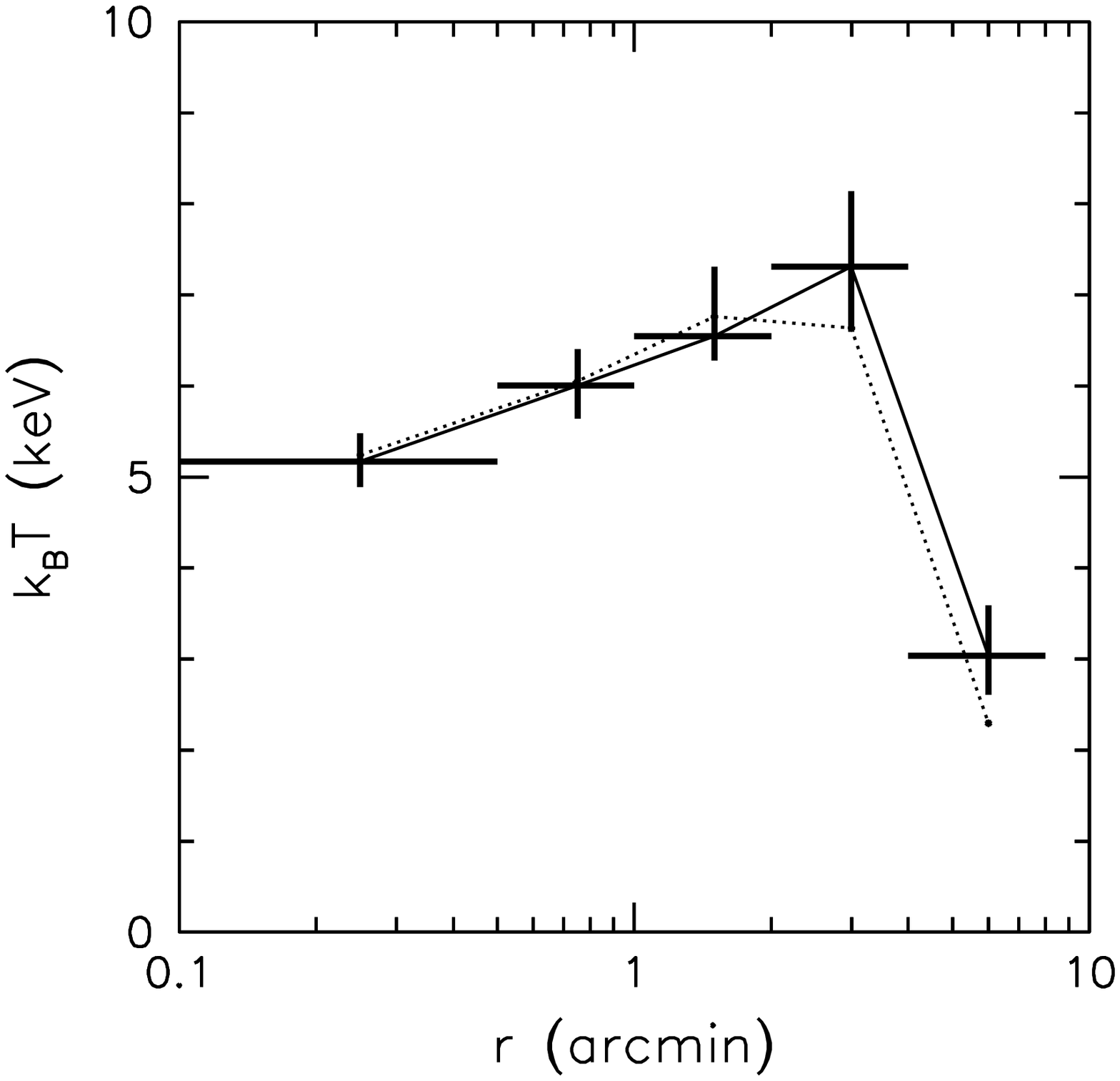}
%{plots/0478_xspec_mo-promewapo_closedrsp.ktprof.ps}
%\includegraphics[width=8.4cm]
%{plots/0478_xspec_mo-promewa_closedrsp.ktprof.ps}
\end{center}
\figcaption{Temperature profiles for RXCJ0307.0$-$2840 from spectral fits with
(solid lines) and
without (dashed lines) a geometrical deprojection.
%``projct$*$mekal$*$wabs'' (right) models in the 0.4--10~keV energy
%band for the MOS1 (dotted lines), MOS2 (dashed lines), pn
%(dash-dotted lines ) and combined (solid lines) data. Solid lines
%connect the temperatures obtained from the combined data.
\label{f:deprojct}
}
\end{figure}

To test the validity of the results without a geometrical deprojection
in our analysis, we apply the deprojection model provided in 
XSPEC (``projct'') to study the
deprojection effect for RXCJ0307.0$-$2840. 
This model performs a three dimension to two dimension 
projection of shells 
onto annuli. It is assumed 
that the inner boundary is specified by the outer boundary of the next 
annulus in. In the ``projct'' model, for each shell in a combined
fit to all annuli spectra simutaneously, the contribution of each
ellipsoidal shell to each annulus is determined and the spectral fitting 
results are then determined. In this fitting the outer shells are not 
affected by the emission from the inner shells.
Similar work has
been described by Pizzolato et al. (2003). Figure~\ref{f:deprojct} presents
the temperature profiles from the spectral fit with and 
without projected modeling in
the 0.4--10~keV energy band. The temperature gradient becomes slightly
more significant when the geometrical deprojection effect is taken
into account. The differences are, however, within the error bars and
can thus be neglected. The relatively small effect of the deprojection
is due to the steep surface brightness profiles of clusters
which strongly reduce the influence of the emission from the outer 
shells on the observed spectra of the central regions. Therefore,
the application of deprojection gives really significant 
improvement only if the count statistics is very high (e.g. 
Matsushita et al. 2002).

Systematic differences in the temperature profiles caused by the
inclusion of the 0.4--1~keV energy band in the spectral analysis are
not the same among the clusters. Cluster RXCJ1131.9$-$1955 is not
affected at all, RXCJ0014.3$-$3022, RXCJ0307.0$-$2840 and
RXCJ0528.9$-$3927 are 
affected in the center, while RXCJ0043.4$-$2037 and
RXCJ0232.2$-$4420 are affected in the outskirts. Since the instrumental
setup used to observe this sample is the same, it hardly is an
instrumental artifact. However, more detailed analyses are needed in
order to distinguish between the Galactic and extragalactic origin of this
component in the outskirts of some of the clusters in our sample
(e.g. Finoguenov et al. 2003).

\subsection{Modeling RXCJ0307.0$-$2840}

We use RXCJ0307.0$-$2840 as an illustrative example to demonstrate the
accuracy of measurements of the total gravitating cluster mass and the
gas mass fraction attainable with the XMM-Newton observations of the
REFLEX-DXL-like clusters. Similar analyses of the REFLEX-DXL clusters
are in progress. RXCJ0307.0$-$2840 is also very bright, but at
a slightly lower redshift than the selection range for the REFLEX-DXL
clusters.

\subsubsection{Gas distribution}

\begin{figure}
\begin{center}
\includegraphics[width=8.4cm]{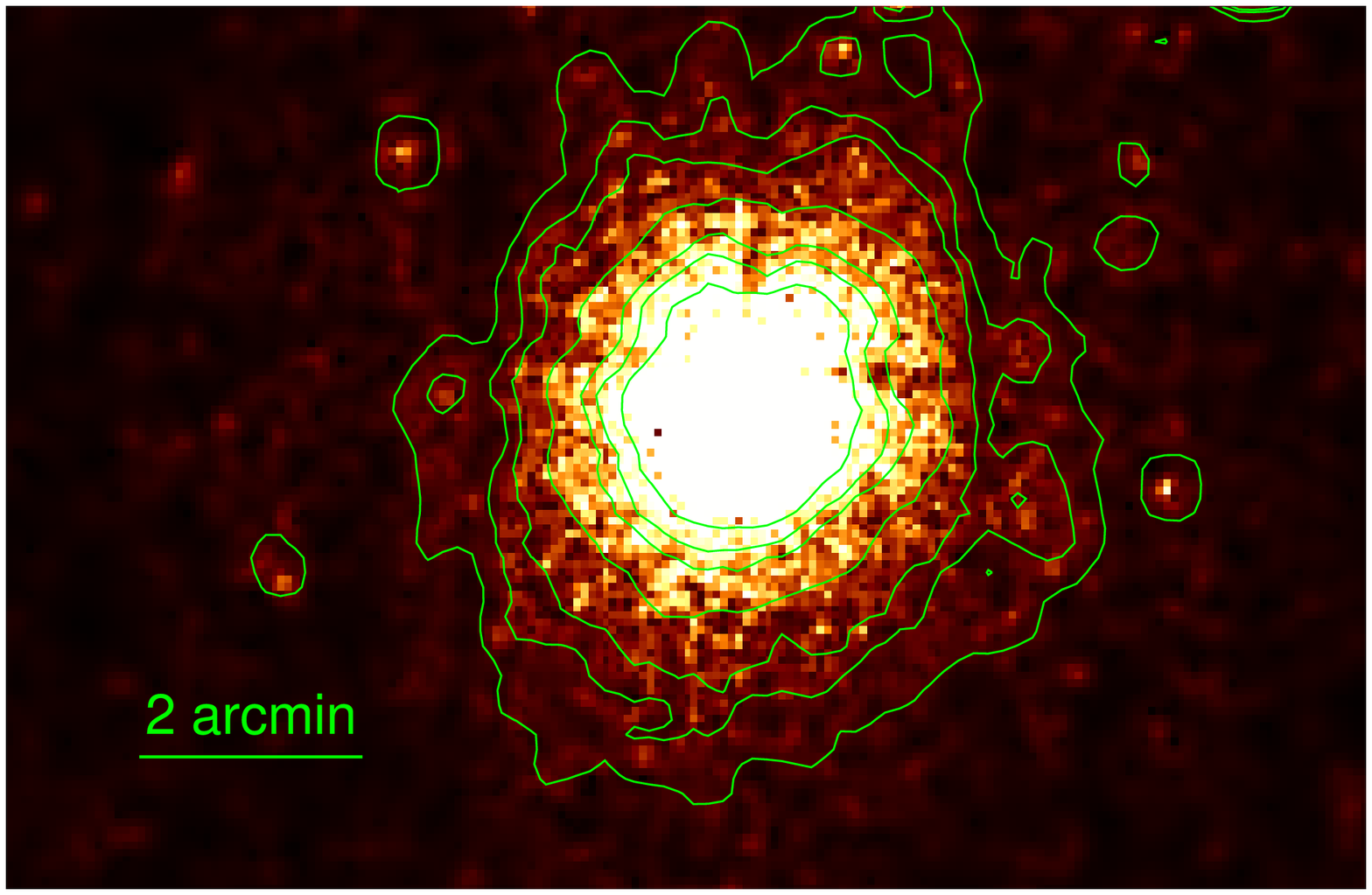}
\end{center}
\figcaption{Merged image from the three instruments for 
RXCJ0307.0$-$2840 in the 0.5--2~keV band.  Superposed contours suggest
a quite regular morphology.
\label{f:CLi_05_20}
}
\end{figure}

The regularity of the photon distribution shown in
Fig.~\ref{f:CLi_05_20} suggests that there are no large substructures
in RXCJ0307.0$-$2840. We thus assume a radially symmetric gas
distribution.  In order to get the actual gas distribution we directly
deproject the data from the spectroscopic analysis to get the gas mass
profile (cf. Fabian et al 1981; Kriss et al. 1983). We divide the
rings used for the temperature determinations into small subrings
and fit the normalization for each subring separately, fixing the
temperature profile to the values obtained from the above spectral
analysis, the metallicities to $Z=0.2Z_\odot$, the redshifts to
$z_{\rm opt}$ as given in Table~\ref{t:cl_all}, and the Galactic
absorption to $n_{\rm H}$ as given in Table~\ref{t:primarytab}. 
We use a constant metallicity here since we did not detect 
a significant variation of metallicity with radius within the error
limits of our analysis.
In the soft band, the X-ray emission is almost independent of the
temperature (Fabricant et al. 1980).  The gas mass in each spherical
shell is proportional to the square root of the integral emission,
which can be calculated from the normalization of the spectroscopic
analysis (e.g. Vikhlinin et al. 1999).

\goodbreak
\goodbreak

\begin{figure}
\begin{center}
\includegraphics[width=8.4cm]{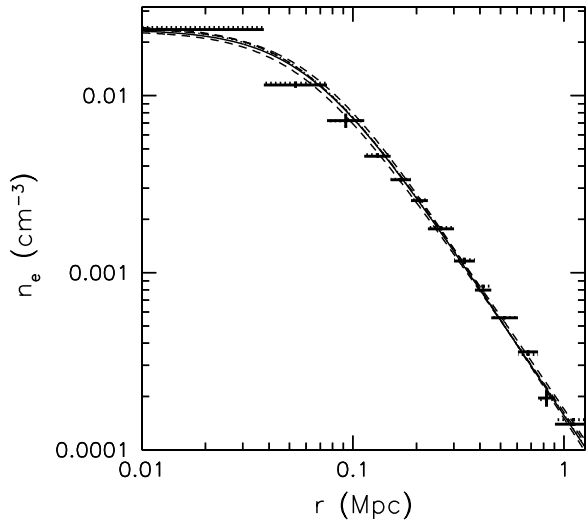}
\end{center}
\figcaption{Measured electron number density profiles of RXCJ0307.0$-$2840
fitted in the 0.4--10~keV band (dotted lines) and 1--10~keV band (solid lines),
respectively.  The corresponding curves present the best fits using the
standard $\beta$ model with the confidence intervals (dashed curves).
\label{f:eleden}
}
\end{figure}

In Fig.~\ref{f:eleden} we show the electron number density profiles for
RXCJ0307.0$-$2840. Here, 1 arcmin~=~0.240~Mpc at $z=0.2578$.  For the
fits we use the standard $\beta$ model

\begin{equation}
n_{\rm e} (r)=
        n_{\rm e0} \left [1 + \left (\frac{r}{r_{\rm c}} \right )^2
        \right ]^{- \frac{3 \beta}{2}}~, 
\label{e:beta}
\end{equation}

with the core radius $r_{\rm c}$ and the shape parameter $\beta$.  The
parameters from the best $\chi^2$ fits are listed in
Table~\ref{t:fit}. The energy band used for the spectral fits does not
affect the normalization, which corresponds to the electron number
density profile (cf. Fig.~\ref{f:eleden}).

\subsubsection{Temperature distribution}

\begin{figure}
\begin{center}
\includegraphics[width=8.4cm]{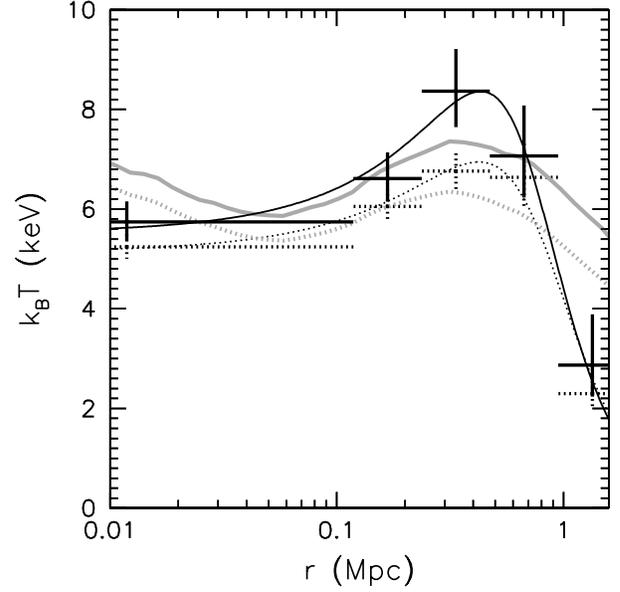}
\end{center}
\figcaption{ Measured temperatures of RXCJ0307.0$-$2840 fitted in the 0.4--10~keV
band (dotted lines) and 1--10~keV band (solid lines). The
corresponding curves show the best fits using the NFW profile (grey) 
and Eq.(\ref{e:tprof}) (black).
\label{f:T_standard}
}
\end{figure}

The precise estimate of the temperature structure greatly contributes
to a reliable mass distribution. The temperature profile of 
RXCJ0307.0$-$2840 drops towards the center due to cooling.  We
found that the parameterization
\begin{equation}
k_{\rm B}T(r)=\frac{1}{A r^2+Br+C }
\label{e:tprof}
\end{equation}
fits the measured temperature profiles quite well.
Fig.~\ref{f:T_standard} presents the best $\chi ^2$ fit for
RXCJ0307.0$-$2840 with the parameters given in Table~\ref{t:fit}. The
temperature profile in the outermost regions can be fitted by a
polytropic model, $k_{\rm B}T(r)=k_{\rm B}T_{0}(n_{\rm e}/n_{\rm e0})^{\gamma -1}$
(e.g. Finoguenov et al. 2001b).  In order for the system 
to be convectionally stable, 
the value of $\gamma$ should not exceed 5/3. Using the
results of the spectral analysis in the 0.4--10~keV (1--10~keV) band
we obtain $\gamma=1.59$ ($\gamma=1.46$), which fulfills the stability
criteria.

\subsubsection{Mass distribution}
\label{s:mass} 

We assume the intracluster gas to be in hydrostatic equilibrium with
the underlying gravitational potential dominated by the dark matter
component.  For the cosmological constant $\Lambda = 0$ we have
\begin{equation}
\frac{1}{\mu m_{\rm p} n_{\rm e}}\frac{d(n_{\rm e} k_{\rm B} T)}{dr}=
  -\frac{GM_{\rm DM}(r)}{r^2}~, 
\label{e:hyd}
\end{equation}
where $n_{\rm e}$ and $k_{\rm B}T$ are the electron number density and
temperature distributions, respectively, and $\mu=0.62$ is the mean
molecular weight per hydrogen atom (e.g. Zakamska \& Narayan 2003).

Analytic models of the gas density and temperature profiles can be
easily combined with Eq.(\ref{e:hyd}) to obtain the mass profile:
\begin{equation}
M_{\rm } (<r)= \frac{r^2}{\mu m_{\rm p} G}k_{\rm B}T(r) \left [
\frac{3 \beta r}{r_{\rm c}^2 + r^2} + (2Ar +B)k_{\rm B}T(r) \right ].
\label{e:mass}
\end{equation}

\vspace*{1cm}

We determine the mass distribution by using our temperature model and
the $\beta$ gas density model.  We use a Mont-Carlo simulation to
calculate the error bars.  
%We take the approximate expression
%$\Delta_{\rm c}(z)=18\pi^2+82[\Omega_{\rm m}(z)-1]-39[\Omega_{\rm
%m}(z)-1]^2$ as the average overdensity of a dark matter halo 
%within the virial radius $r_{\rm vir}$ with respect to the redshift
%dependent critical density $\rho_{\rm crit}(z)$ for a flat
%universe (Bryan \& Norman 1998).

\begin{figure}
\begin{center}
\includegraphics[width=8.4cm]{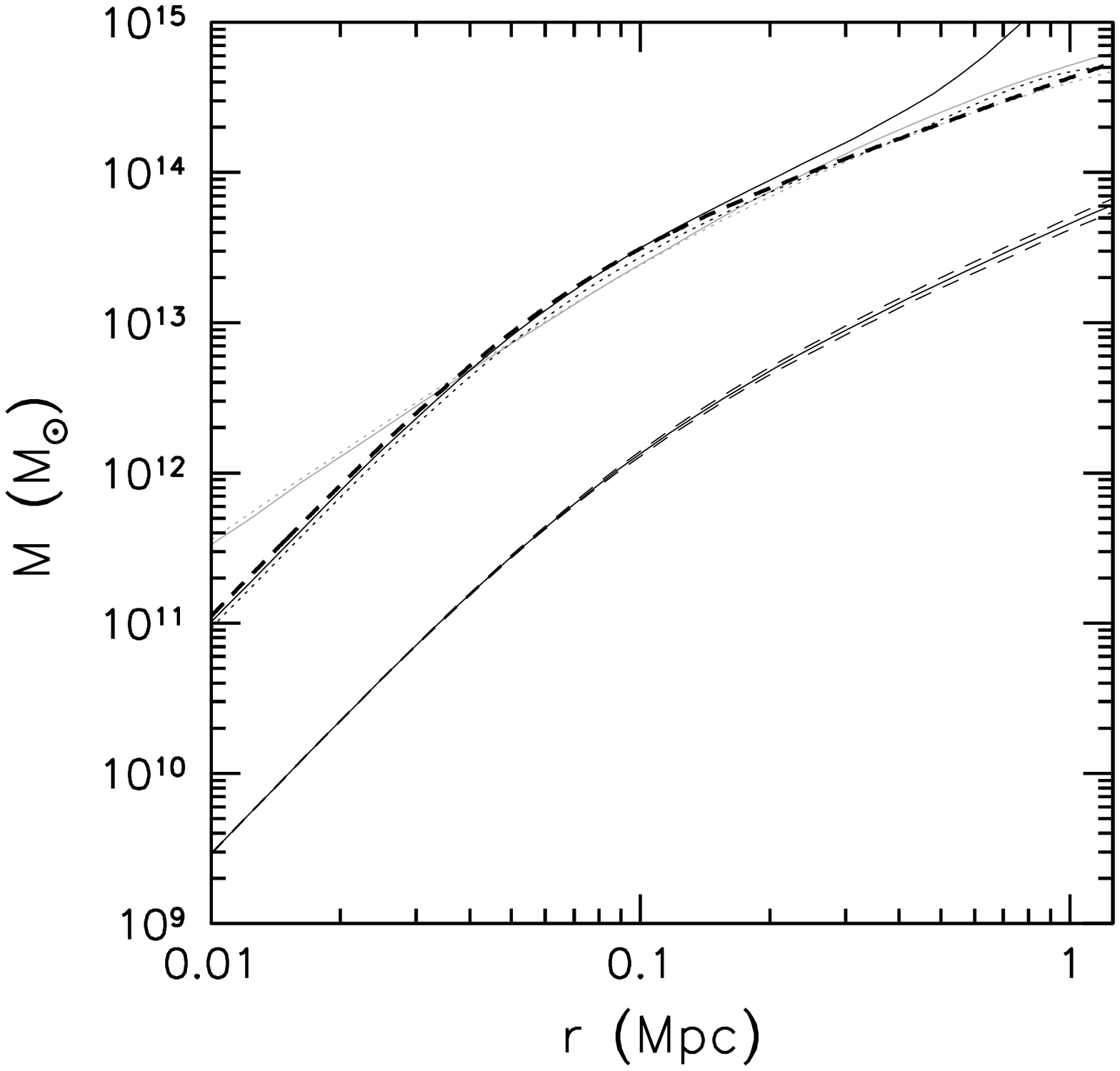}
\end{center}
\figcaption{Mass profiles (top) 
for RXCJ0307.0$-$2840 based on the temperature measurements fitted in
the 0.4--10~keV band (dotted curves) and 1--10~keV band (solid curves)
using the NFW model (grey) and Eq.(\ref{e:mass}) 
(black). An additional dashed curve presents the mass profile under the
assumption of isothermality. The solid curve (bottom) with the
confidence intervals (dashed curves) presents the gas mass
distribution.
\label{f:profile}
}
\end{figure}

Masses measured by the strong gravitational lensing are sometimes
found to be larger compared to the measured masses based on X-rays
(B\"ohringer et al. 2000; Wu 2000; Wu et al. 1998).  High spatially
resolved temperature profiles could help to resolve this discrepancy.
In order to test the effects of temperature gradients we compare
the mass estimates obtained under the assumptions of isothermality
using the global temperature as measured in the 
$0.5<r<4^\prime$ region fitted in the 1--10~keV band, 
and of non-isothermality.  The mass and gas mass
profiles are plotted in Fig.~\ref{f:profile}.  Under the assumption of
hydrostatic equilibrium and isothermality, the virial radius and
the total gravitational cluster mass are 2.14~Mpc and $8.8 \times
10^{14}$ ${\rm M_{\odot}}$, respectively.

\subsubsection{Modeling a NFW mass distribution}
\label{s:nfw} 

Navarro et al. (1997; NFW) described a universal density profile from
numerical simulations in hierarchical clustering scenarios,
%2
\begin{equation}
\rho_{\rm DM}(r)=\frac{\delta_{\rm crit}\rho_{\rm crit}}
                 {(r/r_s)(1+r/r_s)^2}~, 
\label{e:nfw}
\end{equation}
where $\delta_{\rm crit}$ and $r_s$ are the characteristic density and
scale of the halo, respectively, and $\rho_{\rm crit}$, the critical
density of the universe at the cosmic epoch $z$.  $\delta_{\rm crit}$
is related to the concentration parameter of a dark halo $c=r_{\rm
vir}/r_s$ by
%2
\begin{equation}
\delta_{\rm crit}=\frac{200}{3}\frac{c^3}{\ln(1+c)-c/(1+c)}~.
\label{e:nfwc}
\end{equation}
We fit the observational temperature profile to obtain the parameters
$\rho_s=\delta_{\rm crit}\rho_{\rm crit}$ and $r_s$ if we assume that
the hot gas is in hydrostatic equilibrium with the dark matter. The
former is well fitted by a standard $\beta$ profile. The parameters of
the best fit of the NFW profile are presented in Table~\ref{t:fit}.
The virial radius and virial mass estimates are smaller than the estimates
under the assumption of isothermality.

The NFW model describes the mass and gas mass fraction in the
outer region well. Due to the cuspy NFW profile in the cluster center, the
temperature fit based on the NFW model is higher than the
observations. As a result, the gas mass fraction becomes lower in the
center. But for this small central region we can 
not resolve the temperature
structure well enough to perfectly recover the dark matter mass profile
at the small radii.

\goodbreak
\goodbreak

\goodbreak
\goodbreak

\goodbreak
\goodbreak

\goodbreak
\goodbreak

\subsubsection{Modified hydrostatic equilibrium with $\Lambda$}

To be consistent with our background cosmological model with $\Lambda
\ne 0$ we should expect a second-order modification of the equation 
of hydrostatic equilibrium in the form

\begin{equation}
\frac{1}{\mu m_{\rm p} n_{\rm e}}\frac{d(n_{\rm e} k_{\rm B} T)}{dr}=
  -\frac{GM_{\rm DM}(r)}{r^2}+\frac{\Lambda c^2}{3}r~.
\label{e:hydlambda}
\end{equation}

%In Fig.~\ref{f:lam} we present both the fraction $\Lambda c^2
%r^3/(3GM_{\rm DM}(r))$ caused by the $\Lambda$ term and the relative
%error $\Delta M_{\rm DM}(r)/M_{\rm DM}(r)$, where $M_{\rm DM}$ and
%$\Delta M_{\rm DM}(r)$ are obtained from Sect.~\ref{s:mass} and
%Sect.~\ref{s:nfw}. 
The effect of a non-zero $\Lambda$
is smaller than one percent and can thus be neglected compared to the relative
error in our mass
estimations. Sussman \& Hernandez (2003) also point out a small effect of
$\Lambda$ on virialized structures, and that it could be significant only
in the linear regime on large scales of $r \sim 30$~Mpc.

%\begin{center}
%\includegraphics[width=8.4cm]{plots/lambda_4.ps}\\
%\end{center}
%\figcaption{Fraction 
%$\Lambda c^2 r^3/(3GM_{\rm DM}(r))$ (bottom) and the relative error
%$\Delta M_{\rm DM}(r)/M_{\rm DM}(r)$ (top) based on the temperature
%measurements fitted in the 0.4--10~keV band (dotted curves) and
%1--10~keV band (solid curves) using the NFW model (grey) and
%Eq.(\ref{e:mass}) (black), where $M_{\rm DM}$ and $\Delta M_{\rm
%DM}(r)$ are obtained from Sect.~\ref{s:mass} and Sect.~\ref{s:nfw} .
%\label{f:lam}
%}

\subsubsection{Gas mass fraction distribution}

\begin{figure}
\begin{center}
\includegraphics[width=8.4cm]
{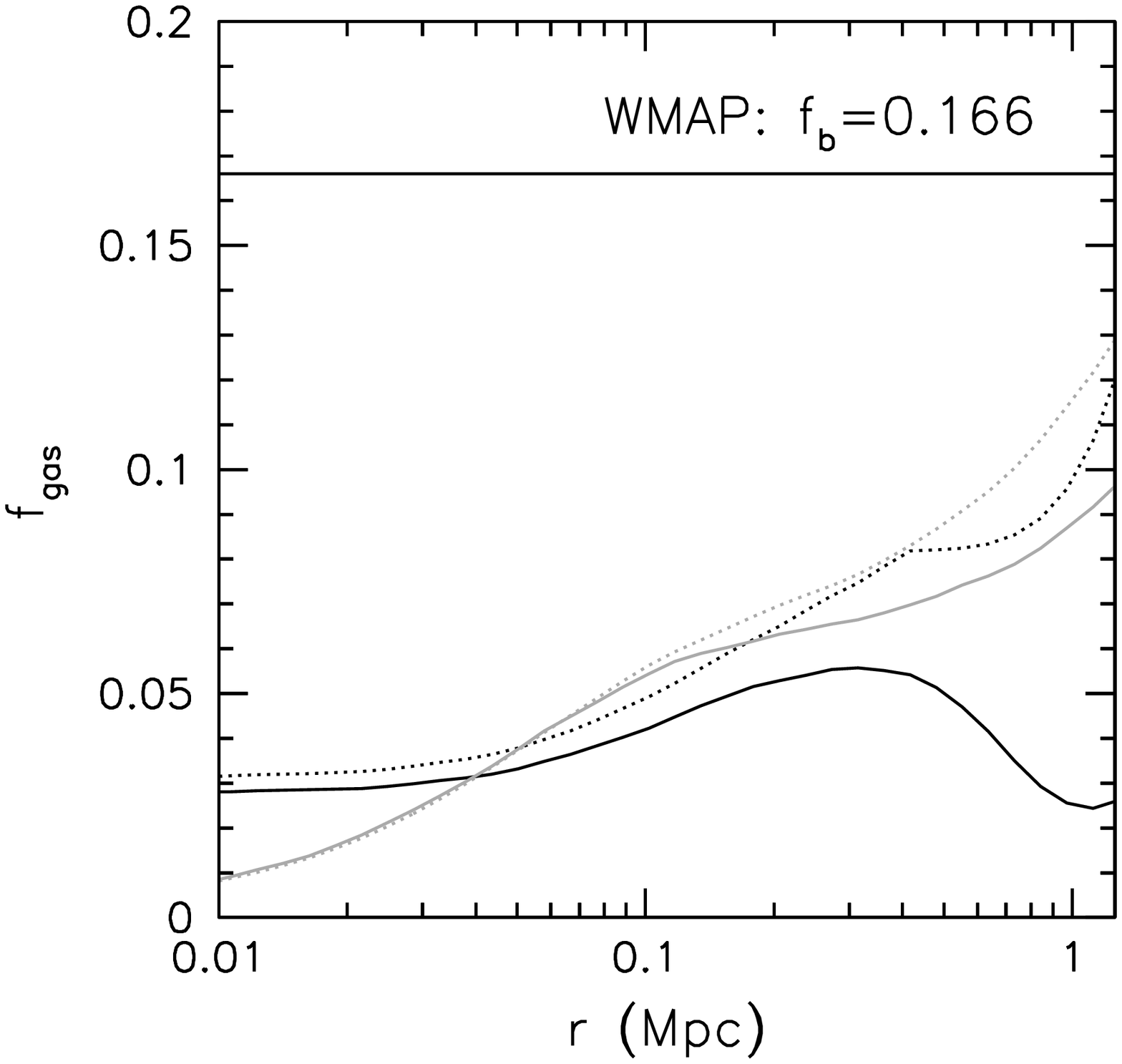}\\
\end{center}
\figcaption{Gas mass fraction distributions 
based on the temperature measurements fitted in the 0.4--10~keV band
(dotted curves) and 1--10~keV band (solid curves) using the NFW model
(grey) and Eq.(\ref{e:mass}) (black).
\label{f:fgasprofile}
}
\end{figure}

The distribution of the gas mass fraction is obtained according to the
definition $f_{\rm gas}(r)=M_{\rm gas}(r)/M_{\rm DM}(r)$.  Since the
derived gas mass is not completely unrelated to the derived total
mass, we calculate the error bars of the gas fraction, 
$\Delta f_{\rm gas}=\sqrt{(\Delta M_{\rm gas} M_{\rm DM})^2 + 
(M_{\rm gas} \Delta
M_{\rm DM})^2}/M^2_{\rm DM}$.  The profile of the gas mass fraction based
on the NFW modeling is steeper in the central region as seen in
Fig.~\ref{f:fgasprofile}.  In both mass modelings the gas fractions
increase with radius, and range from $f_{\rm gas}=0.035\pm0.012$ to
$0.138\pm 0.026$ in the outermost regions ($r_{\rm out}= 1.441$~Mpc)
(see Table~\ref{t:fit}). 

These results are in good agreement with the measurements of Allen et
al. (2002) based on Chandra observations of seven clusters yielding
$f_{\rm gas}\sim$ 0.105--0.138, and with the measurements of Sanderson
et al. (2003) based on ASCA GIS \& SIS and ROSAT/PSPC observations of 66
clusters yielding $f_{\rm gas}=0.13\pm 0.01~h_{70}^{-3/2}$. Our value
is below the universal baryon fraction obtained with the
recent WMAP measurement $f_{\rm b}=\Omega_{\rm b}/\Omega_{\rm
m}=0.166$, where $\Omega_{\rm b}~ h^2=0.0224$ and $\Omega_{\rm m}~
h^2=0.135$ (Spergel et al. 2003).  This reassures the estimate of the
mass distribution.

\goodbreak
\goodbreak

We also make use of the $8<r<10^\prime$ region to obtain an upper
limit of the gas mass in this shell applying the model
``mekal$+$wabs'' without residual background subtraction.  The upper
limit of the total gas mass within 10 arcmin (2.37~Mpc) is $2.49
\times 10^{14} M_{\odot}$.  Based on the above mass modeling described
by Eq.(\ref{e:mass}), the upper limit of the gas fraction within 10
arcmin is 0.49. This unreasonably high value confirms that the
background dominates in this region.

\goodbreak
  \begin{table*} { \begin{center} \footnotesize
  {\renewcommand{\arraystretch}{1.3} \caption[]{Parameters of each
  model from the best $\chi^2$ fits. $r_{\rm out}=1.441$~Mpc is the
  outermost region where we can measure these parameters. $r_{500}$
  and $r_{\rm vir}$ are measured from the data.}  \label{t:fit}}
  \begin{tabular}{llrr}
\hline   
\hline     
 Model        &  parameter                     &   0.4--10~keV     & 1--10~keV \\
\hline  
 $\beta$      & $r_{\rm c}$ (Mpc)  
            & $0.061\pm 0.004$   &  $0.062  \pm 0.004 $\\
              & $n_{\rm e0}$ ($10^{-2}{\rm cm}^{-3}$)
 & $2.43 \pm 0.03$&$ 2.37 \pm0.04 $\\
              & $\beta$                        & $0.60 \pm 0.02$   & $0.60 \pm 0.02$ \\
Eq.(\ref{e:tprof})& $A$ $({\rm~Mpc}^{-2} {\rm~keV}^{-1})
$ & $ 0.283\pm 0.046 $ & $ 0.331\pm 0.082$  \\
              & $B$ $({\rm~Mpc}^{-1} {\rm~keV}^{-1})
$ & $ -0.240\pm 0.028 $ & $ -0.285\pm 0.052$ \\
              & $C$ $({\rm~keV}^{-1})
$ & $ 0.195\pm 0.004$ & $0.181\pm 0.007$\\
NFW& $r_s$ (Mpc)                            &  $ 0.267\pm 0.028$ & $0.306\pm 0.034$\\
      & $\rho_s$ $(10^{15}{\rm M_{\odot}Mpc^{-3}})
      $ & $ 2.2 \pm 0.3$  & $1.9 \pm 0.5$ \\
      & $c $                                   &  $ 5.81\pm0.59 $ & $ 5.56 \pm 0.68 $\\
      & $M_{500}$ $(10^{14}{\rm M_{\odot}})
$  &  $ 4.10\pm 0.37$ & $ 5.13 \pm 0.85$\\
      & $r_{500}$ (Mpc)                        &  $1.038$           & $1.116$ \\
      & $M_{\rm vir}$ $(10^{14}{\rm M_{\odot}})
$ &  $ 5.52 \pm 0.49 $ & $ 7.13\pm 1.18$ \\
      & $r_{\rm vir}$ (Mpc)                    &  $1.550$           & $1.701$ \\
      & $f_{\rm gas}(r<r_{\rm vir})$           &  $ 0.144\pm 0.028$ &  $ 0.124\pm 0.036$\\
      & $M_{r_{\rm out}}$ $(10^{14}{\rm M_{\odot}})
$& $5.17\pm 0.46$  & $ 6.30\pm1.04 $\\
      & $f_{\rm gas}(r<r_{\rm out})$              & $ 0.138\pm 0.026$  & $ 0.112\pm 0.032$\\
Eq.(\ref{e:mass}) &  $M_{500}$ $(10^{14}{\rm M_{\odot}})
$     & $ 6.86\pm 0.01$ &  $5.17 \pm 0.85$\\
      & $r_{500}$ (Mpc)                        & $1.231$           &  $1.118$\\ 
      & $M_{r_{\rm out}}$ $(10^{14}{\rm M_{\odot}})
$& $ 6.91\pm 0.01  $ & $ 16.46 \pm 4.69 $\\
      & $f_{\rm gas}(r<r_{\rm out})$              & $ 0.103 \pm 0.010 $ & $0.035    \pm 0.012$\\
\hline   
\hline  
  \end{tabular}
  \end{center}
\hspace*{0.3cm}{\footnotesize}

}
  \end{table*}
\goodbreak

\section{Summary and Conclusions}
\label{s:conclusion}

We studied eight clusters of the REFLEX-DXL sample, selected from the
REFLEX cluster survey at redshifts around $z\sim 0.3$, and one supplementary
cluster at redshift $z=0.2578$. 
The data are from the MOS1, MOS2 and pn detectors of
XMM-Newton.  The consistent results from the three detectors, obtained
by excluding the energies below 1~keV, give a good confidence in the
applied method and provide tight constraints on the ICM parameters
like the temperature, metallicity, and redshift.

Some of the clusters have been previously studied with ASCA, ROSAT and
Chandra.  The Chandra measurement of the temperature of
RXCJ0658.5$-$5556 includes mainly the central $r<3 ^\prime$ region,
where the temperature is high. We measure the global temperatures over
a larger radial range $0.5<r<4^\prime$ and $r<8^\prime$, respectively.

The parameter which best characterizes cluster mass, and which is
most relevant for studies of the LSS and cosmology, is the hot
temperature of the bulk of the ICM. To avoid contamination by a
possible central cooling core and by a possible soft excess or
residual calibration uncertainties, we excluded the central $r< 0.5
^\prime$ region and the softest part of the X-ray spectrum ($< 1$~keV)
yielding reliable temperatures. These are the global temperatures
that will be used in Paper\,I to derive the X-ray temperature function
for this sample.

We obtained the spatially resolved X-ray temperature profiles for each
cluster.  For the determination of temperature profiles, the good
statistics of the data allowed us to derive temperature values in five
radial bins. In the inner regions an accuracy of better than 10--20 \%
can still be reached while the errors increase in the outer or
outermost two bins.  The temperature varies as a function of radius by
factors of 1.5--2.  The intracluster gas is cool in the center of
RXCJ0232.2$-$4420, RXCJ0307.0$-$2840, RXCJ0528.9$-$3927, and
RXCJ1131.9$-$1955. No significant cooling gas is found with
temperatures below 2~keV.  In the outer region, the temperature drops
at different levels.  The differences of the temperature profiles in
the cluster center may reveal that some clusters have relaxed cooling
cores (but not all) and to some degree that we see the effect of
non-gravitational processes.  In this respect, it is remarkable that
cooling cores are not only found in clusters with symmetric and
regular X-ray images which might suggest a relaxed dynamical state,
but also in the elongated, very disturbed cluster RXCJ1131.9$-$1955.
 
To study RXCJ0307.0$-$2840 in detail, we find a model which fits the
complex temperature profile of this cluster quite well.

The mass distribution of this cluster, based on the precise
measurements of the distributions of the temperature and gas density,
is similar to the mass distribution obtained under the assumption of
isothermality within the region we can measure.  We investigated the
gas mass fraction of RXCJ0307.0$-$2840 and found an increasing gas
mass fraction as a function of radius, which is typical for most
clusters.  In the outermost region of the cluster, it is below the
value of the universal baryon fraction.  The uncertainty of the gas
fraction is mainly caused by the temperature measurement. Therefore, a
reliable determination of the temperature profile is a key point to
obtain the precise estimates of both the mass and the gas mass
fraction. It plays an important role in the M-T scaling relation.

\begin{acknowledgements}
  
The XMM-Newton project is
supported by the Bundesministerium f\"ur Bildung und Forschung,
Deutschen Zentrum f\"ur Luft und Raumfahrt (BMBF/DLR), the Max-Planck
Society and the Haidenhaim-Stiftung. We acknowledge Jacqueline
Bergeron, PI of the XMM-Newton observation of the CDFS, and Martin
Turner, PI of the XMM-Newton observation of RXJ0658.5-5556.  We
acknowledge Steve Sembay who kindly provides us the software to
generate the rmf for MOS and Wolfgang Pietsch, Michael Freyberg,
Frank Haberl and Ulrich G. Briel providing useful suggestions.  YYZ
acknowledges receiving the International Max-Planck Research School
Fellowship.  AF acknowledges receiving the Max-Planck-Gesellschaft
Fellowship.  PS acknowledges support under the DLR grant
No.\,50\,OR\,9708\,35.  YYZ thanks Linda Pittroff for careful reading
the manuscript and useful suggestions.
\end{acknowledgements}
%\end{document}

\bibliographystyle{aabib99}

\end{document}